\documentclass[journal]{IEEEtran}
\ifCLASSINFOpdf
\else
\fi
\usepackage{graphicx}
\usepackage{cite}
\usepackage{picinpar}
\usepackage{amsmath}
\usepackage{amssymb}
\usepackage{url}
\usepackage[latin1]{inputenc}
\usepackage{colortbl}
\usepackage{soul}
\usepackage{multirow}
\usepackage{pifont}
\usepackage{color}
\usepackage{alltt}
\usepackage[hidelinks]{hyperref}
\usepackage{enumerate}
\usepackage{siunitx}
\usepackage{breakurl}
\usepackage{epstopdf}
\usepackage{abraces}
\usepackage{pbox}
\usepackage{bm} 
\usepackage{booktabs} 
\usepackage{threeparttable} 
\usepackage{float} 
\usepackage{subfigure}
\usepackage{empheq}
\usepackage{cases}
\usepackage{hyperref}

\newtheorem{definition}{Definition}
\newtheorem{lemma}{Lemma}
\newtheorem{theorem}{Theorem}

\newtheorem{assumption}{Assumption}

\begin{document}
%
\title{Composite Adaptive Control for Anti-Unwinding Attitude Maneuvers: An Exponential Stability Result Without Persistent Excitation}
%
%
%

	\author{
		\vskip 1em
		{Xiaodong Shao, Qinglei Hu, \emph{Senior Member, IEEE}, Daochun Li, Yang Shi, \emph{Fellow, IEEE}, and Bowen Yi
		}

	\thanks{
		
		
		{		
		X. Shao and D. Li are with the School of Aeronautic Science and Engineering, Beihang University, Beijing 100191, China (e-mail: xdshao\_sasee@buaa.edu.cn; lidc@buaa.edu.cn).
		
		Q. Hu is with the school of Automation Science and Electrical Engineering, Beihang University, Beijing 100191, China (e-mail: huql\_buaa@buaa.edu.cn). 
		
		Y. Shi is with the Department of Mechanical Engineering, University of Victoria, Victoria, BC V8N 3P6, Canada (e-mail: yshi@uvic.ca). 
		
		B. Yi is with the Australian Center for Field Robotics, The University of Sydney, Sydney NSW 2006, Australia (e-mail: bowen.yi@sydney.edu.au).
		}
	}
	}

%
%

\markboth{}
{Shell \MakeLowercase{\textit{\textit{et al.}}}: Bare Demo of IEEEtran.cls for IEEE Journals}
%



\maketitle


\begin{abstract}
This paper provides an exponential stability result for the adaptive anti-unwinding attitude tracking control problem of a rigid body with uncertain but constant inertia parameters, without requiring the satisfaction of persistent excitation (PE) condition. Specifically, a composite immersion and invariance (I\&I) adaptive controller is derived by integrating a prediction-error-driven learning law into the dynamically scaled I\&I adaptive control framework, wherein we modify the scaling factor so that the algorithm design does not involve any dynamic gains. To avoid the unwinding problem, a barrier function is introduced as the attitude error function, along with the tactful establishment of two crucial algebra properties for exponential stability analysis. The regressor filtering method is adopted in combination with the dynamic regressor extension and mixing (DREM) procedure to acquire the prediction error using only easily obtainable signals. In particular, aiding by a constructive liner time-varying filter, the scalar regressor of DREM is extended to generate a new exciting counterpart. In this way, the derived controller is shown to permit closed-loop exponential stability without PE, in the sense that both output-tracking and parameter estimation errors exponentially converge to zero. Further, the composite learning law is augmented with a power term to achieve synchronized finite/fixed-time parameter convergence. Numerical simulations are performed to verify the theoretical findings.
\end{abstract}

\begin{IEEEkeywords}
Exponential stability, immersion and invariance, anti-unwinding, composite adaptive control, attitude tracking.
\end{IEEEkeywords}

%
\IEEEpeerreviewmaketitle

\section{Introduction}
%
%
%
%

\IEEEPARstart{R}{igid}-body attitude control has attracted ever-increasing research attention over the past decades. The interest is not only motivated by atmospheric and space flight applications, but also arises from some other applications ranging from underwater/ground vehicles to rigid robotic systems \cite{chaturvedi2011rigid}. The attitude dynamics for many of the aforementioned rigid-body mechanical systems usually tend to be nonlinear and even uncertain, making the design of attitude controllers that can achieve satisfactory control performance a challenging task. Several elegant solutions to the attitude control problem have been reported in the literature since the early 1990s, such as proportional-derivative (PD) plus feed-forward control (known as ``PD+'' control) \cite{wen1991attitude,arjun2020uniform}, inverse optimal control \cite{luo2005inverse}, model predictive control \cite{li2017continuous}, sliding mode control \cite{huang2018adaptive}, geometric control \cite{lee2017geometric}, etc., as well as a sophisticated combination of these methods. This paper specifically considers the three-axis attitude tracking control of a fully actuated rigid body with full-state feedback. The unit quaternion, a four-parameter representation that is computationally efficient and can globally describe attitude without singularity, is chosen to parameterize the rigid body orientation. 

Due to the redundancy of the unit-quaternion representation, its state space, denoted as $\mathbb{S}^{3}$ (the set of unit-magnitude vectors in $\mathbb{R}^{4}$), is a double cover of $SO(3)$ and, consequently, every physical orientation corresponds to two (antipodal) unit quaternions. It implies that the unit-quaternion tracking error accounts for two equilibria, only one of which is considered \textit{a priori} while the other is left unstable in most existing related works (see, indicatively, \cite{wen1991attitude,luo2005inverse} and references therein). In this manner, certain control actions may cause the rigid body to rotate unnecessarily through a large angle slew or even almost a full revolution to reach the reference trajectory, even if the initial configuration is very close to it, giving rise to the unwinding problem. This phenomenon was discussed at length in \cite{bhat2000topological} from the perspective of lifts of paths and vector fields from $SO(3)$ to $\mathbb{S}^{3}$. In practical applications, such unwinding inevitably leads to inefficient usage of momentum-management devices or fuel of the rigid body \cite{bhat1998continuous}. Kristiansen \textit{et al}. \cite{kristiansen2009satellite} presented a discontinuous backstepping control scheme based on the attitude error function $(1-|q_{e4}|)$ ($q_{e4}$ refers to the scalar part of the unit-quaternion tracking error) to avoid the unwinding behaviors. Later, Mayhew \textit{et al.} \cite{mayhew2011quaternion} developed a hybrid control approach incorporating hysteresis-based switching to mitigate the sensitivity of discontinuous feedbacks presented, for example, in \cite{kristiansen2009satellite} to measurement noises. In \cite{hu2015spacecraft,hu2017unified}, a class of modified attitude error functions were proposed, whereby smooth control laws were obtained for anti-unwinding attitude maneuvers. Costic \textit{et al.} \cite{costic2001quaternion} introduced a barrier function to design anti-unwinding adaptive controllers in both full-state and output feedback fashions. An alternative barrier function for unwinding avoidance was provided in \cite{hu2019anti}. Recently, Dong \textit{et al.} \cite{dong2021anti} derived a sliding mode controller, in which a specially-constructed sliding surface was designed to avoid unwinding. 

Notwithstanding the fact that the methods suggested in the aforementioned works indeed avoid unwinding, most of them have only been shown to deliver uniform asymptotic stability for the closed-loop dynamics, while requiring exact knowledge of the inertia parameters of the rigid body. However, in practice, the rigid body inertia properties are usually uncertain due to, for example, fuel consumption, payload variation, appendage deployment, etc., which limit the opportunities for practical adoption of those anti-unwinding attitude controllers. Adaptive control has been extensively studied as an effective tool to handle parameter uncertainties \cite{tao2003adaptive}. In particular, the successful application of adaptive control theory to the rigid-body attitude tracking problem has been enabled by the crucial fact that the governing attitude-tracking dynamics permits affine regression of inertia-related terms \cite{costic2001quaternion,thakur2015adaptive,filipe2015adaptive,bisheban2020geometric}. A typical feature of these adaptive control solutions is that they are based upon the classical certainty-equivalence (CE) principle. In theoretical terms, the CE-based adaptive controllers can recover the ideal closed-loop performance obtained in the deterministic case wherein the inertia matrix is fully available, only when the persistent excitation (PE) hypothesis holds such that the parameter estimates rapidly converge to their true values. Unfortunately, the PE condition is rarely satisfied in practical scenarios, such as setpoint regulation, close-range proximity operations, etc., hence the performance resulted by the CE-based adaptive controllers is often seen to be poor relative to the uncertainty-free control case. Periodic reference signals can be injected to induce PE \cite{filipe2015adaptive}, but at the cost of unnecessary rotations and energy consumption. To surmount this drawback, Astolfi and Ortega \cite{astolfi2003immersion} departed from the CE principle and instead proposed a new paradigm, known as immersion and invariance (I\&I) adaptive control, yielding a non-CE adaptive controller that possesses better performance. In the I\&I adaptive control framework, a function that satisfies a certain partial differential equation (PDE) is introduced, in combination with the learning term from the adaption law, to generate the parameter estimates. But, for general multi-input systems, there usually exist no solutions to the involved PDEs. This is the so-called ``integrability obstacle''. The main approaches to overcome this problem can be classified into two categories: regressor filtering method \cite{seo2008high,seo2009non} and dynamic scaling method \cite{karagiannis2009dynamic,yang2017dynamically,wen2018dynamic}. Note that the CE and non-CE adaptive control methods mentioned above can only achieve asymptotic convergence of the tracking errors, rather than closed-loop asymptotic stability, making them fragile in the absence of sufficient excitation; moreover, they both require the PE condition for parameter convergence. Indeed, it is sometimes necessary to identify the inertia parameters of the rigid body, e.g., after docking/capture operations, after the deployment of payloads or solar panels, etc. 

Theoretically, a stronger stability result -- exponential stability -- is beneficial for enhancing the robustness of the closed-loop system and for obtaining fast and high-precision attitude tracking performance. However, to the authors' knowledge, there is no previous work achieving adaptive anti-unwinding attitude tracking with exponential stability guarantee, under the quaternion parameterization. The technical obstacles lie in two aspects. On the one hand, the commonly used quaternion-based attitude error functions can hardly exhibit the algebra properties necessary for proving the closed-loop exponential stability, unless additional restrictions on the attitude evolution or control gains are placed, as done in \cite{wen1991attitude}. On the other hand, the rank deficiency of information matrix necessitates the restrictive PE condition for parameter convergence. Overcoming simultaneously the above two obstacles is the main motivation of this paper. Toward this end, a novel composite I\&I adaptive control scheme is proposed by blending a prediction-error-driven learning law into the dynamically scaled I\&I adaptive control framework. We modify the dynamic scaling factor to an essentially bounded one following the line of \cite{xia2020immersion} so that it does not be practically used in the controller implementation, thus greatly reducing the algorithm complexity. The designed controller is shown to deliver exponential stability for the closed-loop dynamics under the assumption of interval excitation (IE), a condition much weaker than PE, in the sense that both the output-tracking and estimation errors exponentially converge to zero without causing unwinding. Such results are partially aided by the choice of a barrier function as the attitude error function to avoid the unwinding phenomenon, along with the tactful establishment of two algebra properties (see Lemmas \ref{Lemma1} and \ref{Lemma2} in Sec. \ref{secII-D}) for exponential stability analysis. In addition, the regressor filtering approach is adopted in conjunction with the dynamic regressor extension and mixing (DREM) procedure recently developed in \cite{aranovskiy2016performance} to obtain the prediction-error-driven learning law using only easily-obtainable signals. Specifically, by virtue of a linear time-varying (LTV) filter \cite{yi2021almost}, a constructive procedure is proposed for DREM to generate, from a set of scalar linear regressor equations (LREs), a set of new scalar LREs in which the new regressor satisfies the PE condition, if the original regressor is of IE. Benefiting from the new LREs, the extended I\&I adaptive law ensures global exponential convergence of the parameter estimation errors under the assumption of IE, not PE. Besides, the composite adaptive law is further generalized to achieve synchronized finite/fixed-time parameter convergence via a slight modification.

The remainder of the paper is organized as follows. Section \ref{secII} presents the problem formulation and mathematical preliminaries. The composite I\&I adaptive controller design is detailed in Section \ref{secIII}, along with rigorous theoretical analyses. In Section \ref{secIV}, some implications are further revealed for the key features and extendibility of the designed controller. Simulation results, which illustrate the effectiveness of the proposed attitude control method, are provided in Section \ref{secV}. Finally, this paper is wrapped-up with concluding remarks.  
%

\section{Problem Formulation and Preliminaries} \label{secII}  

Throughout the paper, vectors and matrices are written in boldface, $\mathbb{R}^{n}$ denotes the $n$-dimensional Euclidean space, and $\mathbb{R}^{m\times n}$ denotes the vector space of $m\times n$ real matrices. $\textbf{I}_{n}$ is the $n\times n$ identity matrix. For a matrix or vector $\bm{A}$, $A_{ij}$ (respectively, $A_{i}$) denotes its $(i,j)$-th (respectively, $i$-th) entry, while $\|\bm{A}\|$ denotes either the Euclidean vector norm or the induced matrix norm. We further write $|\cdot|$ for the absolute value of a scalar and $\text{sign}(\cdot)$ for the standard sign function. The notation $\bm{S}(\cdot):\mathbb{R}^{3} \to \mathbb{R}^{3\times3}$ is a cross product operator such that $\bm{S}(\bm{x})\bm{y}=\bm{x}\times\bm{y}$ for any vectors $\bm{x},\bm{y}\in\mathbb{R}^{3}$. In addition, the set of unit quaternions is given by $\mathbb{Q}_{u}=\{\bm{q}=[\bm{q}_{v}^{\top},q_{4}]^{\top}\in\mathbb{R}^{3}\times\mathbb{R}\mid \bm{q}_{v}^{\top}\bm{q}_{v}+q_{4}^{2}=1\}$. 

\subsection{Rigid-Body Attitude Dynamics} \label{secII-A}

This paper is concerned with the attitude tracking of a rigid body, and three coordinate frames are involved in describing the attitude motion of the rigid body, i.e., the inertial frame $\mathcal{F}_{\mathcal{I}}$, the body-fixed frame $\mathcal{F}_{\mathcal{B}}$, and the reference frame $\mathcal{F}_{\mathcal{R}}$. In $\mathcal{F}_{\mathcal{B}}$, we denote by $\bm{q}=[\bm{q}_{v}^{\top},q_{4}]^{\top}\in\mathbb{Q}_{u}$ and $\bm{\omega}\in\mathbb{R}^{3}$ the inertial attitude and angular velocity of the rigid body, respectively. The kinematic and dynamic equations of the rigid body can be described as \cite{seo2008high}
\begin{equation}
\label{eq1}
\dot{\bm{q}}_{v}=\dfrac{1}{2}(\bm{S}(\bm{q}_{v})+q_{4}\textbf{I}_{3})\bm{\omega},~\dot{q}_{4}=-\dfrac{1}{2}\bm{q}_{v}^{\top}\bm{\omega}
\end{equation}
\begin{equation}
\label{eq2}
\bm{J}\dot{\bm{\omega}}=-\bm{S}(\bm{\omega})\bm{J}\bm{\omega}+\bm{u}
\end{equation}
where $\bm{J}=\bm{J}^{\top}\in\mathbb{R}^{3\times3}$ is the positive definite inertia matrix of the rigid body, and $\bm{u}\in\mathbb{R}^{3}$ is the control torque. Further, define $\bm{q}_{r}=[\bm{q}_{rv}^{\top},q_{r4}]^{\top}\in\mathbb{Q}_{u}$ as the reference quaternion that specifies the rotation of $\mathcal{F}_{\mathcal{R}}$ from $\mathcal{F}_{\mathcal{I}}$, and let $\bm{\omega}_{r}\in\mathbb{R}^{3}$ be the reference angular velocity expressed in $\mathcal{F}_{\mathcal{R}}$. To formulate the attitude tracking problem, an error quaternion $\bm{q}_{e}=[\bm{q}_{ev}^{\top},q_{e4}]^{\top}\in\mathbb{Q}_{u}$ is introduced to describe the relative attitude of $\mathcal{F}_{\mathcal{B}}$ w.r.t. $\mathcal{F}_{\mathcal{R}}$. According to the multiplication rule of quaternions, $\bm{q}_{e}$ is calculated as follows:
\begin{equation}
\label{eq3}
\bm{q}_{e}=\bm{q}_{r}^{-1}\odot\bm{q}=\left[
\begin{matrix}
q_{r4}\bm{q}_{v}-q_{4}\bm{q}_{rv}+\bm{S}(\bm{q}_{v})\bm{q}_{rv} \\
q_{r4}q_{4}+\bm{q}_{rv}^{\top}\bm{q}_{v}
\end{matrix}\right] 
\end{equation}
where $\bm{q}_{r}^{-1}$ is the inverse of $\bm{q}_{r}$, and ``$\odot$'' is the quaternion multiplication operator. The corresponding angular velocity error is defined as $\bm{\omega}_{e}=\bm{\omega}-\bm{C}\bm{\omega}_{r}$, where the rotation matrix $\bm{C}$ from $\mathcal{F}_{\mathcal{R}}$ to $\mathcal{F}_{\mathcal{B}}$ is given by 
\begin{equation}
\label{eq4}
\bm{C}=(q_{e4}^{2}-\bm{q}_{ev}^{\top}\bm{q}_{ev})\textbf{I}_{3}+2\bm{q}_{ev}\bm{q}_{ev}^{\top}-2q_{e4}\bm{S}(\bm{q}_{ev})
\end{equation}
As is well known, $\bm{C}$ satisfies the following two conditions: $\|\bm{C}\|=1$ and $\dot{\bm{C}}=-\bm{S}(\bm{\omega}_{e})\bm{C}$. Now, the open-loop tracking error dynamics are expressed as \cite{costic2001quaternion}
\begin{equation}
\label{eq5}
\dot{\bm{q}}_{ev}=\dfrac{1}{2}(\bm{S}(\bm{q}_{ev})+q_{e4}\textbf{I}_{3})\bm{\omega}_{e},~\dot{q}_{e4}=-\dfrac{1}{2}\bm{q}_{ev}^{\top}\bm{\omega}_{e}
\end{equation}
\begin{equation}
\label{eq6}
\bm{J}\dot{\bm{\omega}}_{e}=-\bm{S}(\bm{\omega})\bm{J}\bm{\omega}+\bm{J}(\bm{S}(\bm{\omega})\bm{\Omega}-\bar{\bm{\Omega}})+\bm{u}
\end{equation}
where $\bm{\Omega}=\bm{C}\bm{\omega}_{r}$ and $\bar{\bm{\Omega}}=\bm{C}\dot{\bm{\omega}}_{r}$ are defined for brevity.

\begin{assumption}
	\label{Assumption1}
	The inertia matrix $\bm{J}$ is diagonal and constant, but otherwise unknown.	
\end{assumption}

\begin{assumption}
	\label{Assumption2}
	The reference angular velocity $\bm{\omega}_{r}$ is bounded and at least $\mathcal{C}^2$ continuous, and its time derivatives up to order two, i.e., $\dot{\bm{\omega}}_{r}$ and $\ddot{\bm{\omega}}_{r}$, are bounded.	
\end{assumption}

\subsection{Affine Regression}

To design an adaptive attitude controller for the open-loop attitude tracking error dynamics described by \eqref{eq5} and \eqref{eq6}, we first perform a linear regression. To this end, a linear regression operator $\bm{L}[\cdot]:\mathbb{R}^{3}\to\mathbb{R}^{3\times6}$ is introduced such that, for any vector $\bm{x}\in\mathbb{R}^{3}$, there always has $\bm{J}\bm{x}=\bm{L}[\bm{x}]\bm{\theta}$, where $\bm{\theta}=[J_{11},J_{22},J_{33},J_{23},J_{13},J_{12}]^{\top}$ is the unknown parameter vector, and $\bm{L}[\bm{x}]$ has the following form
\begin{equation}
\label{eq7}
\bm{L}[\bm{x}]=\left[
\begin{matrix}
x_{1} & 0     & 0     & 0     & x_{3} & x_{2}\\
0     & x_{2} & 0     & x_{3} & 0     & x_{1}\\
0     & 0     & x_{3} & x_{2} & x_{1} & 0
\end{matrix}\right] 
\end{equation}

A logarithmic barrier function $V_{q}$ is introduced to serve as the attitude error function (AEF) for anti-unwinding attitude tracking of the rigid body \cite{hu2019anti}
\begin{equation}
\label{eq8}
V_{q}=-\alpha\ln q_{e4}^{2},~\forall\alpha>0
\end{equation}
which equals to zero only when $q_{e4}=\pm1$ and tends to infinity as $q_{e4}\to0$. The latter implies that for an initial condition satisfying $\bm{q}_{e}(0)\in\mathbb{Q}_{a}$, if the attitude controller $\bm{u}$ is properly designed such that $V_{q}\in\mathcal{L}_{\infty}$, then the attitude tracking error $\bm{q}_{e}$ would evolve strictly in $\mathbb{Q}_{a}$ without unwinding. As a matter of fact, the barrier function $V_{q}$ imposes a permissible set $\mathbb{Q}_{a}=\{\bm{q}_{e}\in\mathbb{Q}_{u}\mid q_{e4}\neq0\}$ for the attitude tracking error $\bm{q}_{e}$. To facilitate the controller design, we further define a filtered tracking error, denoted by $\bm{s}\in\mathbb{R}^{3}$, as follows:
\begin{equation}
\label{eq9}
\bm{s} \doteq \bm{\omega}_{e}+\Lambda\bm{q}_{ev}
\end{equation}
where $\Lambda=\beta/\text{sign}(q_{e4})$ with $\beta>0$ being a design constant. Although $\Lambda$ contains $\text{sign}(q_{e4})$, it is a constant (i.e., $\Lambda=\beta$ or $\Lambda=-\beta$) and has the same sign as $q_{e4}(0)$, under the condition that $q_{e4}(0)\neq0$ and the rotation angle is less than $180^{\circ}$. This condition is actually equivalent to $\bm{q}_{e}(t)\in\mathbb{Q}_{a}$ $\forall t\geq0$. 

From a control perspective, the \textit{target dynamics} desired to be recovered from the uncertainty case are set to
\begin{equation}
\label{eq10}
\dot{\bm{\omega}}_{e}=-k_{p}\bm{s}-\bm{\xi}-\Lambda\dot{\bm{q}}_{ev}
\end{equation}
where $k_{p}>0$ is a constant gain, and $\bm{\xi}=\bm{q}_{ev}/q_{e4}$ (Gibbs vector) is introduced to cancel the cross-coupling term that will appear in the time derivative of $V_{q}$. In view of this, the open-loop dynamics \eqref{eq6} is rewritten as
\begin{align}
\dot{\bm{\omega}}_{e}=&-k_{p}\bm{s}-\bm{\xi}-\Lambda\dot{\bm{q}}_{ev}+\bm{J}^{-1}[\bm{u} \nonumber\\
&\underbrace{-\bm{S}(\bm{\omega})\bm{J}\bm{\omega}
	+\bm{J}(\bm{S}(\bm{\omega})\bm{\Omega}-\bar{\bm{\Omega}}+k_{p}\bm{s}+\bm{\xi}+\Lambda\dot{\bm{q}}_{ev})}_{\bm{\Phi}(\cdot)\bm{\theta}}] \nonumber\\
=& -k_{p}\bm{s}-\bm{\xi}-\Lambda\dot{\bm{q}}_{ev}+\bm{J}^{-1}\left(\bm{u}+\bm{\Phi}(\cdot)\bm{\theta}\right) \label{eq11}
\end{align}
where $\bm{\Phi}(\cdot)\in\mathbb{R}^{3\times6}$ is a known regressor matrix given by
\begin{equation}
\label{eq12}
\bm{\Phi}(\cdot)=-\bm{S}(\bm{\omega})\bm{L}[\bm{\omega}]+\bm{L}[\bm{S}(\bm{\omega})\bm{\Omega}-\bar{\bm{\Omega}}+k_{p}\bm{s}+\bm{\xi}+\Lambda\dot{\bm{q}}_{ev}]
\end{equation}
\subsection{Control Objective} \label{secII-C}

Most existing results on adaptive attitude tracking control can only achieve asymptotic convergence of the system states, rather than closed-loop asymptotic stability, making them fragile in the absence of sufficient excitation. Moreover, from the perspective of practical controller implementation, a stronger stability result, i.e., exponential stability, is more desirable for achieving robust, fast and high-precision attitude tracking. On the other hand, since the rotations around the same axis by Euler angles $0^{\circ}$ and $360^{\circ}$ physically represent the same attitude, the rigid body can certainly track a reference attitude trajectory through performing a relative rotation no larger than $180^{\circ}$. Note, however, that the state space $\mathbb{S}^{3}$ of the unit quaternions is a double cover of $SO(3)$, and there are two antipodal equilibria (i.e., $\bm{q}_{e}=[\bm{0}^{\top},1]^{\top}$ and $\bm{q}_{e}=[\bm{0}^{\top},-1]^{\top}$) for the attitude tracking maneuvers, which may give rise to the so-called unwinding phenomenon. In practical applications, such a phenomenon would lead to unnecessary rotations and inefficient use of momentum-management devices or fuel. Thus, the design of attitude controllers that can achieve exponential attitude tracking without causing unwinding is of both theoretical and practical interest.

Based on the above arguments, the control objective of the paper is to design an adaptive control law $\bm{u}$ for the rigid-body attitude dynamics described by \eqref{eq1} and \eqref{eq2}, such that the resulting closed-loop system is exponentially stable in the sense that $\lim_{t\to\infty}\bm{q}_{e}(t)=[\bm{0}^{\top},\pm 1]^{\top}$, $\lim_{t\to\infty}\bm{\omega}_{e}(t)=\bm{0}$, and $\lim_{t\to\infty}\tilde{\bm{\theta}}(t)=\bm{0}$ ($\tilde{\bm{\theta}}$ refers to the parameter estimation error) at exponential rates without causing unwinding, despite the absence of PE.

\subsection{Definitions and Lemmas} \label{secII-D}

Several definitions and lemmas necessary for the subsequent control design and analysis are introduced in this subsection. First, two fundamental definitions related to signal excitation are presented as follows \cite{tao2003adaptive}: 

\begin{definition} [PE of a Signal]
	\label{Definition1}
	A bounded signal $\bm{X}(t)\in\mathbb{R}^{m\times n}$ is of PE, if there exist constants $t_c>0$ and $\nu>0$ such that $\int_{t}^{t+t_c}\bm{X}(\tau)\bm{X}^{\top}(\tau)d\tau\geq\nu\textbf{I}_{m}$ for all $t\geq 0$.
\end{definition}

\begin{definition} [IE of a Signal]
	\label{Definition2}
	A bounded signal $\bm{X}(t)\in\mathbb{R}^{m\times n}$ is of IE over a time interval $[t_{s},t_{s}+t_{c}]$, if there exist constants $t_{s}\geq0$, $t_{c}>0$ and $\nu>0$ such that $\int_{t_{s}}^{t_{s}+t_{c}}\bm{X}(\tau)\bm{X}^{\top}(\tau)d\tau\geq\nu\textbf{I}_{m}$.
\end{definition}

Next, two pivotal and interesting lemmas are provided, which play central roles in achieving exponential stability of the closed-loop system. We should emphasize that although the two lemmas are established for the error quaternion $\bm{q}_{e}$, they actually hold for any unit quaternion. 

\begin{lemma} 
	\label{Lemma1} 
	For all $\bm{q}_{e}\in\mathbb{Q}_{a}$, the AEF $(-\ln q_{e4}^{2})$ satisfies
	\begin{equation}
	\label{eq13}
	-\ln q_{e4}^{2} \leq \dfrac{1-q_{e4}^2}{|q_{e4}|}
	\end{equation}	
\end{lemma} 

\textit{Proof}: See Appendix A.	

\begin{lemma} 
	\label{Lemma2} 
	Given any scalar $\delta\in(0,1)$, there exist two positive constants $\overline{\alpha}$ and $\underline{\alpha}$ satisfying $\overline{\alpha}\geq-\alpha\ln\delta^{2}/(1-\delta^{2})$ and $\underline{\alpha}\leq1$, such that the following inequality holds for $\delta\leq|q_{e4}|\leq1 $:
	\begin{equation}
	\label{eq14}
	\underline{\alpha}\|\bm{q}_{ev}\|^{2}\leq V_{q}\leq\overline{\alpha}\|\bm{q}_{ev}\|^{2}
	\end{equation}	
\end{lemma} 

\textit{Proof}: See Appendix B.

\section{Adaptive Controller Design} \label{secIII}

In this section, a composite I\&I adaptive control scheme is proposed to achieve the control objective as stated in Sec. \ref{secII-C}. We first construct a solvable PDE by partially reconfiguring the regressor matrix $\bm{\Phi}$, in order to overcome the integrability obstacle that arises in the I\&I adaptive control design. Subsequently, a regressor filtering method is presented to eliminate the need for unmeasurable $\dot{\bm{\omega}}$ in I\&I adaptive augmentation, followed by a constructive DREM procedure for relaxing the restrictive PE assumption for parameter convergence. Then, a dynamically scaled I\&I adaptive controller, augmented with a DREM-based learning law, is designed to achieve anti-unwinding attitude tracking with exponential convergence of the output-tracking and parameter estimation errors, under a weak IE Assumption.

\subsection{Construction of A Solvable PDE}  

For notational concision, let 
\begin{equation}
\label{eq15}
\bm{y}=-\bar{\bm{\Omega}}-k_{p}\bm{\Omega}+k_{p}\Lambda\bm{q}_{ev}+\bm{\xi}-\Lambda\bm{Q}(\bm{q}_{e})\bm{\Omega}
\end{equation}
Decompose $\bm{\Phi}(\cdot)$ in \eqref{eq12} into two parts
\begin{equation}
\label{eq16}
\bm{\Phi}(\cdot)=\bm{\Phi}_{1}(\bm{\omega},\bm{y})+\bm{\Phi}_{2}(\bm{\omega},\bm{\Omega},\bm{q}_{e})
\end{equation}
where we have defined
\begin{equation}
\label{eq17}
\bm{\Phi}_{1}(\bm{\omega},\bm{y})=k_{p}\bm{L}[\bm{\omega}]+\bm{L}[\bm{y}]
\end{equation}
\begin{equation}
\label{eq18}
\bm{\Phi}_{2}(\bm{\omega},\bm{\Omega},\bm{q}_{e})=-\bm{S}(\bm{\omega})\bm{L}[\bm{\omega}]+\bm{L}[\bm{S}(\bm{\omega})\bm{\Omega}]+\Lambda\bm{L}[\bm{Q}(\bm{q}_{e})\bm{\omega}]
\end{equation} 

It is easy to check that $\bm{\Phi}_{1}^{\top}(\bm{\omega},\bm{y})$ is a Jacobian matrix, in other terms, for all $i,j\in\{1,2,3\}$, $\partial \bm{\phi}_{1i}/\partial \omega_{j}=\partial \bm{\phi}_{1j}/\partial \omega_{i}$, where $\bm{\phi}_{1i}$ and $\bm{\phi}_{1j}$ denote the $i$-th and $j$-th columns of $\bm{\Phi}_{1}^{\top}(\bm{\omega},\bm{y})$, respectively. There therefore exists $\bm{\mu}_{1}\in\mathbb{R}^{6}$ such that the following PDE holds: 
\begin{equation}
\label{eq19}
\dfrac{\partial \bm{\mu}_{1}}{\partial \bm{\omega}}=\bm{\Phi}_{1}^{\top}(\bm{\omega},\bm{y})
\end{equation}
As $\bm{y}$ is independent of $\bm{\omega}$, we can obtain a feasible but not unique solution of  \eqref{eq19} as follows: 
\begin{equation} 
\label{eq20}
\bm{\mu}_{1}=\bm{L}^{\top}[\bm{y}]\bm{\omega}+k_{p}\bar{\bm{\omega}}
\end{equation} 
with $\bar{\bm{\omega}}=[ 0.5\omega_{1}^{2},0.5\omega_{2}^{2},0.5\omega_{3}^{2},\omega_{2}\omega_{3},\omega_{1}\omega_{3},\omega_{1}\omega_{2}]^{\top}$.

Unlike $\bm{\Phi}_{1}^{\top}(\bm{\omega},\bm{y})$, the sub-regressor $\bm{\Phi}_{2}^{\top}(\bm{\omega},\cdot)$ (for notational simplicity, the symbol ``$\cdot$'' is used to capture all the other arguments of $\bm{\Phi}_{2}$ except for $\bm{\omega}$) is not a Jacobian matrix, since for all $i,j\in\{1,2,3\}$ except $i=j$, $\partial \bm{\phi}_{2i}/\partial \omega_{j}\neq\partial \bm{\phi}_{2j}/\partial \omega_{i}$, where $\bm{\phi}_{2i}$ and $\bm{\phi}_{2j}$ denote the $i$-th and $j$-th columns of $\bm{\Phi}_{2}^{\top}(\bm{\omega},\cdot)$, respectively. Consequently, there is no $\bm{\mu}_{2}\in\mathbb{R}^{6}$ satisfying the PDE $\partial \bm{\mu}_{2}/\partial \bm{\omega}=\bm{\Phi}_{2}^{\top}(\bm{\omega},\cdot)$. This is commonly known as the ``integrability obstacle'', which prevents the classical I\&I adaptive control method in \cite{astolfi2003immersion} from being directly applied to the problem under study. To overcome such a restrictive obstacle, we construct a solvable PDE by reconfiguring $\bm{\Phi}_{2}(\bm{\omega},\cdot)$ in the following way: 
\begin{equation} 
\label{eq21}
\dfrac{\partial \bm{\mu}_{2}}{\partial \bm{\omega}}=\hat{\bm{\Phi}}_{2}^{\top}(\bm{\omega},\hat{\bm{\omega}},\cdot)\\
\end{equation} 
with $\hat{\bm{\Phi}}_{2}^{\top}(\bm{\omega},\hat{\bm{\omega}},\cdot)$ being $[\bm{\phi}_{21}(\omega_{1},\hat{\omega}_{2},\hat{\omega}_{3},\cdot),\bm{\phi}_{22}(\hat{\omega}_{1},\omega_{2},\hat{\omega}_{3},\cdot),$ $\bm{\phi}_{23}(\hat{\omega}_{1},\hat{\omega}_{2},\omega_{3},\cdot)]$, and $\hat{\bm{\omega}}$ a filter state determined by
\begin{equation} 
\label{eq22}
\dot{\hat{\bm{\omega}}}=-\bar{\bm{y}}-k_{f}\tilde{\bm{\omega}},~\hat{\bm{\omega}}(0)=\bm{\omega}(0)
\end{equation}
where $\bar{\bm{y}}=\bm{y}+k_{p}\bm{\omega}+\bm{S}(\bm{\omega})\bm{\Omega}+\Lambda\bm{Q}(\bm{q}_{e})\bm{\omega}$ and $\tilde{\bm{\omega}}=\hat{\bm{\omega}}-\bm{\omega}$. Noting the fact that $-\bm{S}(\bm{\omega})\bm{J}\bm{\omega}+\bm{J}\bar{\bm{y}}=\bm{\Phi}\bm{\theta}$, and from \eqref{eq2} and \eqref{eq22}, it follows that
\begin{equation} 
\label{eq23}
\dot{\tilde{\bm{\omega}}}=-k_{f}\tilde{\bm{\omega}}-\bm{J}^{-1}(\bm{\Phi}\bm{\theta}+\bm{u})
\end{equation}
A direct solution to \eqref{eq23} is   
\begin{align} 
\bm{\mu}_{2}=&\int_{0}^{\omega_{1}}\hat{\bm{\phi}}_{21}(\tau,\hat{\omega}_{2},\hat{\omega}_{3},\cdot)\text{d}\tau+\int_{0}^{\omega_{2}}\hat{\bm{\phi}}_{22}(\hat{\omega}_{1},\tau,\hat{\omega}_{3},\cdot)\text{d}\tau \nonumber \\
             &+\int_{0}^{\omega_{3}}\hat{\bm{\phi}}_{23}(\hat{\omega}_{1},\hat{\omega}_{2},\tau,\cdot)\text{d}\tau \label{eq24}
\end{align}

Let $\bm{\mu}=\bm{\mu}_{1}+\bm{\mu}_{2}$. Then, from  \eqref{eq16}, \eqref{eq19} and \eqref{eq21}, we get
\begin{equation}
\label{eq25}
\dfrac{\partial \bm{\mu}}{\partial \bm{\omega}}=\bm{\Phi}^{\top}(\cdot)+\bm{\Psi}^{\top}(\cdot)
\end{equation}
where the matrix $\bm{\Psi}(\cdot)\in\mathbb{R}^{3\times6}$ is defined by
\begin{equation}
\label{eq26}
\bm{\Psi}(\cdot)=\hat{\bm{\Phi}}_{2}(\bm{\omega},\hat{\bm{\omega}},\cdot)-\bm{\Phi}_{2}(\bm{\omega},\cdot)
\end{equation}

\subsection{Regressor Filtering and DREM} \label{secIII-B}

From a theoretical viewpoint, extracting information about the unknown inertia parameters from the prediction errors requires measurement of the angular acceleration $\dot{\bm{\omega}}$, which is usually unavailable in practice. Various differentiators such as fixed-point/lag Kalman smoother, high-gain observer, state differentiator, etc., have been advocated to estimate $\dot{\bm{\omega}}$, but they are sensitive to measurement noise and disadvantageous to the rigor of analysis. This motivates us to employ the regressor filtering method to avoid the usage of angular acceleration estimate in parameter adaptation. 

Before proceeding, we rewrite  \eqref{eq2} as $\bm{J}\dot{\bm{\omega}}=\bm{W}\bm{\theta}+\bm{u}$ with $\bm{W}=-\bm{S}(\bm{\omega})\bm{L}[\bm{\omega}]$, and introduce a linear time-invariant (LTI) filter $\mathcal{H}$ of which the transfer function is $\mathcal{H}(s)=\frac{1}{s+a}$, where $a>0$ is the filter time constant. Then, the filtered signals  $\bm{\omega}_{f}$, $\bm{W}_{f}$, and $\bm{u}_{f}\in\mathbb{R}^{3}$ are generated by passing $\bm{\omega}$, $\bm{W}$, and $\bm{u}$, respectively, through $\mathcal{H}$ as follows:
\begin{equation}
\label{eq27}
\bm{\omega}_{f}=\mathcal{H}[\bm{\omega}],~\bm{W}_{f}=\mathcal{H}[\bm{W}],~\bm{u}_{f}=\mathcal{H}[\bm{u}]
\end{equation}
with initial conditions $\bm{\omega}_{f}(0)=\bm{\omega}(0)/a$, $\bm{W}_{f}(0)=\bm{0}$, and $\bm{u}_{f}(0)=\bm{0}$. After a simple calculation, we get
\begin{equation}
\label{eq28}
\dot{\bm{\omega}}_{f}=\bm{J}^{-1}(\bm{W}_{f}\bm{\theta}+\bm{u}_{f})
\end{equation}
Reorganizing  \eqref{eq28} obtains a linear regressor equation (LRE)
\begin{equation}
\label{eq29}
\bm{u}_{f}=(\bm{L}[\dot{\bm{\omega}}_{f}]-\bm{W}_{f})\bm{\theta}=\bm{W}_{a}\bm{\theta}
\end{equation}
As shown above, the filtered input $\bm{u}_{f}$ contains information about the unknown parameter vector $\bm{\theta}$ and, therefore, can act as the measurement for parameter estimation. 

Since $\bm{u}_{f}$, $\dot{\bm{\omega}}_{f}$, and $\bm{W}_{f}$ are computable at every time instance from the LTI filters in  \eqref{eq27}, $\bm{\theta}$ can be extracted from the LRE \eqref{eq29} without involving any unmeasurable signals. A gradient-descent estimator is generally designed to estimate $\bm{\theta}$ and has the form of $\dot{\hat{\bm{\theta}}}=-\Gamma\bm{W}_{a}^{\top}(\bm{W}_{a}\hat{\bm{\theta}}-\bm{u}_{f})=-\Gamma\bm{N}\tilde{\bm{\theta}}$, where $\Gamma>0$ is a constant gain and $\bm{N}=\bm{W}_{a}^{\top}\bm{W}_{a}$ is a symmetric information matrix. It is important to underscore that $\bm{N}$ is at most rank 3 and thus only positive semidefinite. Under this situation, the parameter convergence can be achieved only if the regressor $\bm{W}_{a}^{\top}$ is PE, a condition that is rarely met in practice. The rank deficiency of $\bm{N}$ is the primary cause of parameter convergence requiring PE. To relax the excitation requirement for parameter convergence, the information matrix $\bm{N}$ will be designed to possess full rank under a strictly weak IE assumption, using the recently reported DREM technique \cite{aranovskiy2016performance}. For that, the following is assumed for the degree of excitation in the filtered regressor $\bm{W}_{a}$. 
\begin{assumption}
	\label{Assumption3}
	There exist $t_{s}>0$ and $t_{c}>0$ such that the filtered regressor $\bm{W}_{a}^{\top}$ is of IE over $[t_{s},t_{s}+t_{c}]$. 
\end{assumption}

Following the line of \cite{aranovskiy2016performance}, our first step is to construct an extended LRE (e-LRE) with a square regressor matrix via the Kreisselmeier's regressor extension proposed in \cite{kreisselmeier1977adaptive}, recently revisited in \cite{ortega2020new,yi2020conditions}. The detailed construction proceeds as follows. Pre-multiplying both sides of  \eqref{eq29} by $\bm{W}_{a}^{\top}$ gives
\begin{equation}
\label{eq30}
\bm{W}_{a}^{\top}\bm{u}_{f}=\bm{W}_{a}^{\top}\bm{W}_{a}\bm{\theta}
\end{equation}
to which we apply an LTI filter $\mathcal{K}$ with transfer function $\mathcal{K}(s)=\frac{1}{s+b}$ ($b>0$ is the filter time constant). To be specific, the state-space realization of $\mathcal{K}$ is as follows:
\begin{equation}
\label{eq31}
\dot{\bm{M}}=-b\bm{M}+\bm{W}_{a}^{\top}\bm{u}_{f},~\bm{M}(0)=\bm{0}
\end{equation}
\begin{equation}
\label{eq32}
\dot{\bm{N}}=-b\bm{N}+\bm{W}_{a}^{\top}\bm{W}_{a},~\bm{N}(0)=\bm{0} 
\end{equation}
with the filtered signals $\bm{M}\in\mathbb{R}^{6}$ and $\bm{N}\in\mathbb{R}^{6\times6}$. Solving  \eqref{eq31} and \eqref{eq32} readily arrives at $\dot{\aoverbrace[L1R]{\bm{M}-\bm{N}\bm{\theta}}}=-b(\bm{M}-\bm{N}\bm{\theta})$, from which it is easy to obtain the e-LRE
\begin{equation}
\label{eq33}
\bm{M}=\bm{N}\bm{\theta}
\end{equation}

Next, the regressor \textit{mixing} step is performed to obtain a set of scalar LREs that share the same scalar regressor. We pre-multiply the adjunct matrix, denoted by $\mbox{adj}\{\cdot\}$, of $\bm{N}$ to both sides of the e-LRE \eqref{eq33} to get
\begin{equation}
\label{eq34}
Y_i = \Delta \theta_i, \; i\in\{1,\ldots, 6\}
\end{equation}
with definitions in the following compact form 
\begin{equation}
\label{eq35}
\bm{Y} \doteq k_{\tt I}\text{adj}\{\bm{N}\} \bm{M} \footnote{Based on the Cramer's rule, $Y_{i}$ can be calculated by $Y_{i}=k_{\tt I}\text{det}(\bm{N}_{M,i})$, where $\bm{N}_{M,i}$ is the matrix $\bm{N}$ with its $i$-th column replaced by $\bm{M}$.},~\Delta \doteq k_{\tt I}\det\{\bm{N}\}
\end{equation}
where $k_{\tt I}>1$ is introduced, if necessary, to enhance the regressor signal strength in case of a low level of excitation. But, such a treatment inevitably magnifies the deviation between $Y_{i}$ and $\Delta\theta_{i}$ caused by external disturbances, measurement noises, and some other practical factors, which in turn may decrease the parameter estimation accuracy of the DREM estimator. As such, $k_{\tt I}$ should be judiciously chosen according to the excitation level of the reference trajectory and the mission requirement on identification accuracy. 

Closely inspecting  \eqref{eq32} reveals that the information matrix $\bm{N}$ at any time instance $t$ is calculated by a weighted accumulation (via forward integration) of all incoming data from $0$ up to $t$. In this way, the rank of $\bm{N}$ will be gradually populated over time such that $\bm{N}$ turns to be full rank at a certain moment, if the IE condition as stated in Assumption \ref{Assumption3} holds. In fact, from Assumption \ref{Assumption3}, one can deduce that
\begin{equation}
\label{eq36}
\begin{split}
\bm{N}(t_{s}+t_{c})&\geq\int_{t_{s}}^{t_{s}+t_{c}}e^{-b(t_{s}+t_{c}-\tau)}\bm{W}_{a}^{\top}(\tau)\bm{W}_{a}(\tau)d\tau \\
&\geq e^{-b t_{c}}\int_{t_{s}}^{t_{s}+t_{c}}\bm{W}_{a}^{\top}(\tau)\bm{W}_{a}(\tau)d\tau \\
&\geq\nu e^{-b t_{c}}\textbf{I}_{6}>0
\end{split}
\end{equation}
showing that $\bm{N}(t)$ becomes full rank at $t=t_{s}+t_{c}$ and accordingly $\Delta(t_{s}+t_{c})\geq(\nu e^{-b t_{c}})^{6}$. However, if $\bm{W}_{a}^{\top}$ is only of IE, the value of $\Delta$ will decay to zero with time after the end of IE, due to the exponential forgetting design in \eqref{eq32}. In view of this, the direct use of $\Delta$ in designing the parameter estimator $\dot{\hat{\bm{\theta}}}=-\Gamma\Delta(\Delta\hat{\bm{\theta}}-\bm{Y})$, as done in \cite{aranovskiy2016performance}, may lead to a remarkable decrease in parameter convergence rate over time. To surmount this problem, in the sequel we seek to augment the LREs in  \eqref{eq34} with tactfully introduced filtered states and free terms, with the aim of generating a set of new scalar LREs that share a non-degenerate scalar regressor, under extremely weak Assumption \ref{Assumption3}. 

According to the construction in \cite[Proposition 1]{yi2021almost}, a linear time-varying (LTV) filter is introduced
\begin{equation}
\label{eq37}
\left\lbrace
\begin{aligned} 
\dot{\bm{\chi}} & = \Delta(\bm{Y} - \Delta\bm{\chi})  \\
\dot{\Xi}       & = -\Delta^2\Xi, \;\;  \Xi(0) = 1
\end{aligned}\right. 
\end{equation}
From  \eqref{eq34} and \eqref{eq37}, it is straightforward to get
\begin{equation}
\label{eq38}
\dot{\aoverbrace[L1R]{\bm{\chi} - \bm{\theta}}} = - \Delta^2 (\bm{\chi} - \bm{\theta})
\end{equation}
The solution of the LTV system \eqref{eq38} is given by\footnote{The term $\Xi(t)\Xi(\tau)^{-1}$ is indeed the state transition matrix of the LTV system $\dot{\bm{x}}= -\Delta^2(t)\bm{x}$ from $\tau$ to $t$.}
\begin{equation}
\label{eq39}
\begin{split}
\bm{\chi}(t)-\bm{\theta}&=\exp\left(-\int_0^t \Delta^2(\tau) \rm d\tau\right)(\bm{\chi}(0)-\bm{\theta})\\
&=\Xi(t)(\bm{\chi}(0)- \bm{\theta}) 
\end{split}
\end{equation}
By inserting  \eqref{eq39} into the original LREs \eqref{eq34}, we obtain a set of new scalar LREs as follows:
\begin{equation}
\label{eq40}
\bm{Y}_{\tt N} = \Delta_{\tt N} \bm{\theta}
\end{equation}
where $\bm{Y}_{\tt N}$ and $\Delta_{\tt N}$ are defined, respectively, by
\begin{equation}
\label{eq41}
\bm{Y}_{{\tt N}}(t)\doteq \bm{Y}(t) + k_{\tt N} (\bm{\chi}(t) - \Xi(t)\bm{\chi}(0))
\end{equation}
\begin{equation}
\label{eq42}
\Delta_{\tt N}(t) \doteq \Delta(t) + k_{\tt N} (1 - \Xi(t))
\end{equation}
with $k_{\tt N}>0$ being a design constant.

\begin{lemma} 
	\label{Lemma3}
	The LRE extension based on the LTV filter \eqref{eq37} guarantees that:
	\begin{enumerate}
		\item The LTV filter \eqref{eq37} is internally stable;
		\item The newly obtained scalar regressor $\Delta_{\tt N}$ satisfies $\Delta_{\tt N}(t)\geq0$ on $t\in[0,\infty)$;
		\item If Assumption \ref{Assumption3} holds, then there exists a constant $\hbar>0$ such that $\Delta_{\tt N}(t)>\hbar$ on $t\in[t_{s}+t_{c},\infty)$.
	\end{enumerate}
\end{lemma}

\textit{Proof}: A sketch of the proof is provided here. The interested reader may refer to \cite{yi2021almost} for additional details. 

In view of  \eqref{eq38} and the intuitive fact that $\Delta^{2}(t)\geq0$ $\forall t\geq0$, it can be immediately concluded that $\|\bm{\chi}(t) -\bm{\theta}\|$ is not increasing, showing the internal stability of the LTV filter \eqref{eq37}. Recalling  \eqref{eq32}, we find that the information matrix $\bm{N}$ is a symmetric positive semi-definite, which indicates that $\Delta(t)\geq0$ for all $t\geq0$. Besides, one can easily infer that $(1-\Xi(t))\geq0$. It is then evident from  \eqref{eq42} that $\Delta_{\tt N}(t)\geq0$ on $t\in[0,\infty)$. 

Further, we consider Assumption \ref{Assumption3} holds, that is, $\bm{W}_{a}^{\top}$ is of IE. Given this, from  \eqref{eq36} and its accompanying result $\Delta(t_{s}+t_{c})\geq(\nu e^{-b t_{c}})^{6}$, there exist $t_{\Delta}>0$ and $o>0$ such that
\begin{equation}
\label{eq43}
\int_{t_{s}+t_{c}-t_{\Delta}}^{t_{s}+t_{c}}\Delta(\tau){\rm d}\tau\geq\sqrt{o} \implies \int_{0}^{t}\Delta^{2}(\tau){\rm d}\tau\geq o
\end{equation}
for all $t \geq t_{s}+t_{c}$, from which it is easy to verify the following implications:
\begin{align}
\bm{W}_{a}^{\top}(t) \in \mbox{IE}  \implies & \Delta(t) \in \mbox{IE} \nonumber\\
\implies     &  1 - \Xi(t) \geq 1-e^{-o}>0, \; \forall t \geq t_{s}+t_{c} \nonumber\\
\implies     &  \Delta_{\tt N}(t)>\hbar=k_{\tt N}(1-e^{-o}), \; \forall t \geq t_{s}+t_{c} \nonumber\\
\implies     &  \Delta_{\tt N}(t)\in\mbox{PE}, \; \forall t \geq t_{s}+t_{c} \label{eq44}
\end{align}
where the fact that $\Delta(t)\ge0$ $\forall t\geq0$ has been used in the last two implications. $\hfill \blacksquare$



\subsection{Composite I\&I Adaptive Controller}

At this point, a composite I\&I adaptive controller is designed by applying a dynamically scaled I\&I adaptive control method with parallel combination of a prediction-error-driven (also called DREM-based) learning law.    

Design the control law and the accompanying adaptive laws as
\begin{equation}
\label{eq45}
\bm{u}=-\bm{\Phi}(\hat{\bm{\theta}}+\bm{\zeta})
\end{equation}
\begin{equation}
\label{eq46}
\dot{\hat{\bm{\theta}}}=\underbrace{-\gamma[\dot{\bar{\bm{\mu}}}-(\bm{\Phi}+\bm{\Psi})^{\top}\bar{\bm{y}}]}_{\text{I\&I-based learning law: Part I}}\underbrace{-\gamma\lambda\bm{\epsilon}}_{\text{~~~DREM-based learning law~~~}}
\end{equation}
\begin{equation}
\label{eq47}
\bm{\zeta}=\underbrace{\gamma\bm{\mu}}_{\text{I\&I-based learning law: Part II}}
\end{equation}
where $\gamma,\,\lambda>0$ are constant gains, $\dot{\bar{\bm{\mu}}}=\dot{\bm{\mu}}-(\partial \bm{\mu}/\partial \bm{\omega})\dot{\bm{\omega}}$, and $\bm{\epsilon}\in\mathbb{R}^{6}$ is the prediction error vector given by 
\begin{equation}
\label{eq48}
\bm{\epsilon}=\Delta_{\tt N}(\hat{\bm{\theta}}+\bm{\zeta})-\bm{Y}_{\tt N}
\end{equation}
Actually, the composite term $(\hat{\bm{\theta}}+\bm{\zeta})$ acts as the estimate of $\bm{\theta}$, thus the parameter estimation error is defined as $\tilde{\bm{\theta}}=\hat{\bm{\theta}}+\bm{\zeta}-\bm{\theta}$. From  \eqref{eq40} and \eqref{eq48}, it follows that $\bm{\epsilon}=\Delta_{\tt N}\tilde{\bm{\theta}}$. We underscore that the DREM-based learning law $-\gamma\lambda\bm{\epsilon}$ can extract actual information of $\bm{\theta}$ from the rich historical data.

Inserting  \eqref{eq45} into  \eqref{eq11} yields
\begin{equation}
\label{eq51}
\dot{\bm{\omega}}_{e}= -k_{p}\bm{s}-\bm{\xi}-\Lambda\dot{\bm{q}}_{ev}-\bm{J}^{-1}\bm{\Phi}\tilde{\bm{\theta}}
\end{equation}
Bearing  \eqref{eq2}, \eqref{eq25}, \eqref{eq46}-\eqref{eq48} in mind and recalling the definitions of $\bm{\Phi}$ and $\bar{\bm{y}}$, we have
\begin{align}
\dot{\tilde{\bm{\theta}}}&=\gamma\dfrac{\partial \bm{\mu}}{\partial \bm{\omega}}\dot{\bm{\omega}}+\gamma\dot{\bar{\bm{\mu}}}-\gamma\left[\dot{\bar{\bm{\mu}}}-(\bm{\Phi}+\bm{\Psi})^{\top}\bar{\bm{y}}\right]-\gamma\lambda\bm{\epsilon} \nonumber\\
&=\gamma(\bm{\Phi}+\bm{\Psi})^{\top}\bm{J}^{-1}\left(-\bm{S}(\bm{\omega})\bm{J}\bm{\omega}+\bm{J}\bar{\bm{y}}+\bm{u}\right)-\gamma\lambda\bm{\epsilon} \nonumber\\
&=-\gamma(\bm{\Phi}+\bm{\Psi})^{\top}\bm{J}^{-1}\bm{\Phi}\tilde{\bm{\theta}}-\gamma\lambda\bm{\epsilon} \label{eq52}
\end{align}
In what follows, the dynamic scaling technique is applied to obviate the effect of perturbation $\bm{\Psi}$ on parameter estimation. However, for most I\&I adaptive control methods with dynamic scaling, the lower bound (or minimum eigenvalue) of the unknown parameter needs to be known beforehand; on the other hand, the scaling factor does not involve a damping term in its dynamics and thus grows monotonically and continuously under perturbed and noisy conditions. The latter may result in ``high-gain'' control actions and further cause undesirable transient behavior and robustness degradation of the closed-loop system. To surmount the disadvantages above, inspired by \cite{xia2020immersion}, a naturally bounded scaling factor $R(t)$ that satisfies $R(t)>R_{0}$ $\forall t\geq0$ for some constant $R_{0}>0$ is presented to form the scaled estimation error
\begin{equation}
\label{eq53}
\bm{z}=\dfrac{\tilde{\bm{\theta}}}{R}
\end{equation}
with $R$ having the following form 
\begin{equation}
\label{eq54}
R=\dfrac{\sqrt{J_{\rm m}}}{e^{1/(2J_{\rm m}^{2})}}\cdot e^{\frac{\sqrt{\ln f(r)}}{J_{\rm m}}}
\end{equation}
where $J_{\rm m}$ denotes the minimum eigenvalue of $\bm{J}$, and $f(r)$ is defined as 
\begin{equation}
\label{eq55}
f(r)=f_{\rm m}\tanh(r)+1
\end{equation}
In the above formula, $f_{\rm m}>0$ can be freely chosen to adjust the maximum value of $f(r)$ (noting that $1< f(r)\leq f_{\rm m}+1$), while $r$ is a time-varying scalar satisfying $r(t)>0$ $\forall t\geq0$ and is determined by 
\begin{equation}
\label{eq56}
\dot{r}=\gamma\dfrac{f(r)\sqrt{\ln f(r)}}{\partial f(r)/ \partial r}\|\bm{\Psi}\|^{2},~~r(0)>0
\end{equation}
Taking the time derivative of $\bm{z}$ and noting  \eqref{eq52}-\eqref{eq56} lead to
\begin{equation}
\label{eq57}
\dot{\bm{z}}=-\gamma(\bm{\Phi}+\bm{\Psi})^{\top}\bm{J}^{-1}\bm{\Phi}\bm{z}-\gamma\lambda\Delta_{\tt N}\bm{z}-\dfrac{\gamma}{2J_{\rm m}}\|\bm{\Psi}\|^{2}\bm{z}
\end{equation}

Now, consider a Lyapunov-like function
\begin{equation}
\label{eq58}
V_{z}=\dfrac{1}{2\gamma}\bm{z}^{\top}\bm{z}
\end{equation}
By Young's inequality, we deduce that
\begin{equation}
\label{eq59}
\dot{V}_{z}\leq-\dfrac{J_{\rm m}}{2}\|\bm{J}^{-1}\bm{\Phi}\bm{z}\|^{2}-\lambda\Delta_{\tt N}\|\bm{z}\|^{2}\leq0
\end{equation}
where the fact $\Delta_{\tt N}(t)\geq0$ $\forall t\geq0$ shown in Lemma \ref{Lemma3} has been used. Thus, the equilibrium $\bm{z}=\bm{0}$ of the scaled estimation error dynamics \eqref{eq57} is uniformly globally stable and accordingly uniformly bounded.

The main contributions of this paper are summarized in the following theorem.

\begin{theorem} \label{Theorem1}
	Consider the rigid-body attitude dynamics described by  \eqref{eq1} and \eqref{eq2} under Assumptions \ref{Assumption1} and \ref{Assumption2}. Given the initial conditions $[\bm{q}(0),\bm{\omega}(0)]$ and the reference trajectories $[\bm{q}_{r}(0),\bm{\omega}_{r}(0)]$ satisfying $\bm{q}_{e}(0)\in\mathbb{Q}_{a}$, if $k_{p}$ and $k_{f}$ are chosen as $k_{p}=k_{f}=\kappa (f_{\rm m}+1)$ with $\kappa>0$ some design constant, then the implementation of the control law \eqref{eq45} in conjunction with the parameter estimator defined in  \eqref{eq46} and \eqref{eq47} leads to the following:
	\begin{enumerate}	
		\item All closed-loop trajectories converge asymptotically to an invariant attracting manifold $\mathcal{M}$ given by
		\begin{equation}
		\label{eq60}
		\mathcal{M}\doteq\{\tilde{\bm{\theta}}\in\mathbb{R}^{6}\mid\bm{\Phi}\tilde{\bm{\theta}}=\bm{0}\}
		\end{equation}
		as a result, the uncertain plant dynamics are ultimately immersed into the target dynamics \eqref{eq10};
		\item The output-tracking errors $\bm{q}_{ev}(t)$ and $\bm{\omega}_{e}(t)$ asymptotically converge to zero on $t\in[0,\infty)$, and the unwinding phenomenon is strictly avoided;
		\item If Assumption 3 holds, the origin of the closed-loop system is exponentially stable on $t\in[t_{s}+t_{c},\infty)$, in the sense that $\bm{q}_{ev}(t)$, $\bm{\omega}_{e}(t)$, and $\tilde{\bm{\theta}}$(t) converge to zero exponentially fast on $t\in[t_{s}+t_{c},\infty)$.
	\end{enumerate}  
\end{theorem}

\textit{Proof}: Consider the overall Lyapunov-like function
\begin{equation}
\label{eq61}
V=V_{q}+\dfrac{1}{2}\bm{s}^{\top}\bm{s}+\dfrac{1}{2}\tilde{\bm{\omega}}^{\top}\tilde{\bm{\omega}}+\eta V_{z}
\end{equation}
where $\eta=2(1/\kappa+\varrho)$ with $\varrho>0$ is introduced just for stability analysis. Taking the time derivative of $V$ and noting  \eqref{eq5}, \eqref{eq9}, \eqref{eq23}, \eqref{eq51} and \eqref{eq59}, we have
\begin{align}
\dot{V}\leq&~\dfrac{\bm{q}_{ev}^{\top}}{q_{e4}}\bm{\omega}_{e}+\bm{s}^{\top}\dot{\bm{s}}+\tilde{\bm{\omega}}^{\top}\dot{\tilde{\bm{\omega}}}-\dfrac{\eta J_{\rm m}}{2}\|\bm{J}^{-1}\bm{\Phi}\bm{z}\|^{2}-\eta\lambda\Delta_{\tt N}\|\bm{z}\|^{2} \nonumber\\
\leq&-\dfrac{\beta}{|q_{e4}|}\bm{q}_{ev}^{\top}\bm{q}_{ev}-k_{p}\bm{s}^{\top}\bm{s}-R\bm{s}^{\top}\bm{J}^{-1}\bm{\Phi}\bm{z}-k_{f}\tilde{\bm{\omega}}^{\top}\tilde{\bm{\omega}} \nonumber\\
&+R\tilde{\bm{\omega}}^{\top}\bm{J}^{-1}\bm{\Phi}\bm{z}-\dfrac{\eta J_{\rm m}}{2}\|\bm{J}^{-1}\bm{\Phi}\bm{z}\|^{2}-\eta\lambda\Delta_{\tt N}\|\bm{z}\|^{2} \label{eq62}
\end{align}

By Young's inequality, we have
\begin{subequations} 
	\begin{equation}
	\label{eq63a}
	R\leq\dfrac{\sqrt{J_{\rm m}}}{e^{1/(2J_{\rm m}^{2})}}\cdot e^{\left(\frac{\ln f(r)}{2}+\frac{1}{2J_{\rm m}^{2}}\right)}\leq\sqrt{J_{\rm m}}\sqrt{f(r)} \tag{61a}
	\end{equation}
	\begin{equation}
	\label{eq63b}
	-R\bm{s}^{\top}\bm{J}^{-1}\bm{\Phi}\bm{z}\leq\dfrac{\kappa f(r)}{2}\|\bm{s}\|^{2}+\dfrac{J_{\rm m}}{2\kappa}\|\bm{J}^{-1}\bm{\Phi}\bm{z}\|^{2} \tag{61b}
	\end{equation}
	\begin{equation}
	\label{eq63c}
	R\tilde{\bm{\omega}}^{\top}\bm{J}^{-1}\bm{\Phi}\bm{z}\leq\dfrac{\kappa f(r)}{2}\|\tilde{\bm{\omega}}\|^{2}+\dfrac{J_{\rm m}}{2\kappa}\|\bm{J}^{-1}\bm{\Phi}\bm{z}\|^{2} \tag{61c}
	\end{equation}
\end{subequations}
Using \eqref{eq63a}-\eqref{eq63c} in \eqref{eq62} and further from the fact that $\eta=2(1/\kappa+\varrho)$ and $1< f(r)\leq f_{\rm m}+1$, it follows that
\begin{align}
\dot{V}\leq&-\dfrac{\beta}{|q_{e4}|}\|\bm{q}_{ev}\|^{2}-\dfrac{\kappa(f_{\rm m}+1)}{2}\|\bm{s}\|^{2}-\dfrac{\kappa(f_{\rm m}+1)}{2}\|\tilde{\bm{\omega}}\|^{2} \nonumber\\
&-\varrho     
J_{\rm m}\|\bm{J}^{-1}\bm{\Phi}\bm{z}\|^{2}-\eta\lambda\Delta_{\tt N}\|\bm{z}\|^{2} \label{eq65}
\end{align}

Inspecting \eqref{eq65}, it is found that $\dot{V}(t)\leq0$ for all $t\geq0$, from which we establish the boundedness of $V$ and hence $V_{q}$. The latter, together with the fact that $\bm{q}_{e}(0)\in\mathbb{Q}_{a}$, imply that there exists a positive constant $\delta\in(0,1)$ such that $\bm{q}_{e}$ remains in a compact subset $\mathbb{B}\subset\mathbb{Q}_{a}$, defined by $\mathbb{B}=\{\bm{q}_{e}\in\mathbb{Q}_{u}\mid \delta\leq|q_{e4}|\leq1\}$, for all $t\geq0$; in other words, the set $\mathbb{Q}_{a}$ is positive invariant. Thus, the unwinding phenomenon is strictly avoided during the entire mission. It is then clear that $\bm{\xi}$ in  \eqref{eq10} is of $\mathcal{L}_{\infty}$. By integrating both sides of \eqref{eq65}, we know that $\int_{0}^{\infty}\dot{V}(\tau)\text{d} \tau$ exists and is finite, which in turn implies that $\bm{q}_{ev}$, $\bm{s}$, $\tilde{\bm{\omega}}$, and $\bm{J}^{-1}\bm{\Phi}\bm{z}\in\mathcal{L}_{2}\cap\mathcal{L}_{\infty}$. By the boundedness of $\bm{q}_{ev}$ and $\bm{s}$, we have $\bm{\omega}_{e}\in\mathcal{L}_{\infty}$ from  \eqref{eq9}. Further, invoking Assumption \ref{Assumption2} and the fact that $\bm{\omega}_{e},\,\tilde{\bm{\omega}}\in\mathcal{L}_{\infty}$, one can easily deduce $\bm{\omega}\in\mathcal{L}_{\infty}$, so does $\hat{\bm{\omega}}$. Based on the above argument, it is evident from  \eqref{eq12} that $\bm{\Phi}\in\mathcal{L}_{\infty}$ and from  \eqref{eq21} and \eqref{eq26} that $\bm{\Psi}\in\mathcal{L}_{\infty}$. As the scaling factor $R$ is naturally bounded (see \eqref{eq54}), $\bm{J}^{-1}\bm{\Phi}\bm{z}\in\mathcal{L}_{2}\cap\mathcal{L}_{\infty}$ is actually equivalent to $\bm{J}^{-1}\bm{\Phi}\tilde{\bm{\theta}}\in\mathcal{L}_{2}\cap\mathcal{L}_{\infty}$. In addition, the state and regressor filtering operations $\mathcal{H}[\bm{\omega}]$ and $\mathcal{H}[\bm{W}]$ described by  \eqref{eq27} with bounded inputs generate bounded signals $\bm{\omega}_{f}$ and $\bm{W}_{f}$, respectively, which leads to the boundedness of $\bm{W}_{a}$ and $\bm{N}$. From $\bm{N}\in\mathcal{L}_{\infty}$, we show that $\Delta$ and hence $\Delta_{\tt N}\in\mathcal{L}_{\infty}$. It is now straightforward to check from  \eqref{eq52} that $\dot{\tilde{\bm{\theta}}}\in\mathcal{L}_{\infty}$. By noticing  \eqref{eq11} and \eqref{eq12}, we have $\dot{\bm{\omega}}_{e},\,\dot{\bm{\Phi}}\in\mathcal{L}_{\infty}$. A direct deduction gives the conclusion that $\dot{\bm{q}}_{ev}$, $\dot{\bm{s}}$, $\dot{\tilde{\bm{\omega}}}$, and $\frac{\text{d}}{\text{d} t} (\bm{J}^{-1}\bm{\Phi}\tilde{\bm{\theta}})$ are bounded and, accordingly, uniformly continuous. Then, by applying Barbalat?s lemma, we guarantee that
\begin{equation}
\label{eq66}
\lim_{t\to\infty}[\bm{q}_{ev}(t),\bm{s}(t),\tilde{\bm{\omega}}(t),\bm{J}^{-1}\bm{\Phi}(t)\tilde{\bm{\theta}}(t)]=\bm{0}
\end{equation}

From  \eqref{eq9} and \eqref{eq66}, it follows that $\lim_{t\to\infty}\bm{\omega}_{e}(t)=\bm{0}$. In fact, the convergence condition $\lim_{t\to\infty}\bm{J}^{-1}\bm{\Phi}(t)\tilde{\bm{\theta}}(t)=\bm{0}$ directly contributes to the establishment of an invariant attracting manifold $\mathcal{M}$ as defined in  \eqref{eq60}. In view of this, all the closed-loop trajectories asymptotically converge to $\mathcal{M}$, showing that the uncertain plant dynamics will be immersed into the target dynamics \eqref{eq10}. Consequently, the ideal closed-loop performance obtained in the deterministic case can be ultimately recovered without requiring convergence of the parameter estimates to the corresponding true values.  

Next, we consider Assumption \ref{Assumption3} holds such that $\Delta_{\tt N}(t)>\hbar>0$ for all $t\geq t_{s}+t_{c}$, as stated in Lemma \ref{Lemma3}. Under this condition, let $\bm{Z}\doteq[\bm{q}_{ev}^{\top},\bm{s}^{\top},\tilde{\bm{\omega}}^{\top},\bm{z}^{\top}]^{\top}$. Then, recalling the fact that $|q_{e4}|\geq\delta$ ($\delta\in(0,1)$) and Lemma \ref{Lemma2}, it can be easily shown that $V$ in  \eqref{eq61} is bounded by the following:
\begin{equation}
\label{eq67}
\underline{\varsigma}\|\bm{Z}\|^{2}\leq V(\bm{q}_{ev},\bm{s},\tilde{\bm{\omega}},\bm{z})\leq\overline{\varsigma}\|\bm{Z}\|^{2}
\end{equation}
where $\underline{\varsigma} \doteq \min\{\underline{\alpha},1/2,\eta/2\gamma\}$ and $\overline{\varsigma} \doteq \max\{\overline{\alpha},1/2,\eta/2\gamma\}$. Further, by invoking Lemma \ref{Lemma1},  \eqref{eq65} becomes
\begin{align}
\dot{V}&\leq\beta\ln(q_{e4}^{2})-\dfrac{\kappa(f_{\rm m}+1)}{2}\|\bm{s}\|^{2}-\dfrac{\kappa(f_{\rm m}+1)}{2}\|\tilde{\bm{\omega}}\|^{2} \nonumber \\
       &-\eta\lambda\hbar\|\bm{z}\|^{2}\leq-\varsigma V \label{eq68}
\end{align}
for all $t \geq t_{s}+t_{c}$, where $\varsigma\doteq\min\{\beta,\kappa(f_{\rm m}+1),2\gamma\lambda\hbar\}$. By the comparison lemma, we get
\begin{equation}
\label{eq69}
V(t)\leq V(T_{s})\exp(-\varsigma (t-T_{s})),~~\forall t\geq T_{s}
\end{equation}
where $T_{s}=t_{s}+t_{c}$ is defined hereafter for notational brevity. This implies that $V(t)\to0$ uniformly exponentially fast on $t\in[T_{s},\infty)$, which together with  \eqref{eq67} allow us to conclude the exponential stability of the equilibrium point $\bm{Z}=\bm{0}$ on $t\in[T_{s},\infty)$ by \cite[Theorem 4.10]{khalil2002nonlinear}. Further, from  \eqref{eq9} and the exponential convergence of $\bm{q}_{ev}(t)$ and $\bm{s}(t)$ on $t\in[T_{s},\infty)$, we conclude that $\bm{\omega}_{e}(t)$ converges to zero exponentially fast on $t\in[T_{s},\infty)$, thus completing the proof.  $\hfill \blacksquare$

\section{Implications} \label{secIV}

In light of the foregoing theoretical results, the following implications are summarized in order.

1) The key steps that permit achieving exponential stability for the resulting closed-loop system on $t\in[T_{s},\infty)$ are to deduce the formulas \eqref{eq67}-\eqref{eq69} under Assumption \ref{Assumption3} (a weak excitation), through the judicious introduction of a DREM-based larning term $-\gamma\lambda\bm{\epsilon}$ into the adaptive law \eqref{eq46} and our establishment of Lemmas \ref{Lemma1} and \ref{Lemma2} for $V_{q}$. Strictly speaking, the stability condition shown in this paper can be considered as a local result, since the augmented I\&I adaptive controller itself provides exponential stability on $t\in[T_{s},\infty)$ for the initial conditions satisfying $\bm{q}_{e}(0)\in\mathbb{D}$ ($\mathbb{D}\subseteq\mathbb{B}$), a subset of $\mathbb{Q}_{a}$. Although it is difficult to quantify the interior bound of $\mathbb{D}$, from a qualitative viewpoint, $q_{e4}(0)$ can be chosen as close as possible to zero until the actuators under magnitude limits cannot provide sufficient torques to ensure system stability. Thus, the local result has a quite large domain of attraction.

Some other barrier functions, e.g., $(1-q_{e4}^{2})/q_{e4}^{2}$, may be used as an alternative to $V_{q}$ in \eqref{eq8} for achieving exponential stability and unwinding avoidance. Additionally, it is important to emphasize that the commonly used AEFs, for example, $2(1-q_{e4})$ and $2(1-q_{e4}^{2})$, can hardly display the algebra properties similar to Lemmas \ref{Lemma1} and \ref{Lemma2}, unless some additional restrictions on the attitude convergence or high-gain feedback assumptions are placed. For instance, Wen and Kreutz-Delgado \cite{wen1991attitude} chose $2(1-q_{e4})$ as the AEF and demonstrated exponential stability, but they require $q_{e4}\to+1$ such that $q_{e4}(t)\geq0$ holds after a finite time, whilst laying lower bounds on the control gains to ensure that certain nonlinear terms are adequately dominated in the Lyapunov sense. This is the main technical barrier for most of the existing attitude control methods to obtaining an exponential stability result, even if the inertia matrix $\bm{J}$ is known.

2) The composite I\&I adaptive controller derived in this paper preserves the key features of the I\&I adaptive control methodology via establishment of an invariant attracting manifold $\mathcal{M}$ defined in \eqref{eq60}, while achieving local exponential stability for the closed-loop system on $t\in[T_{s},\infty)$ under a weak IE Assumption. Apart from this, an essentially bounded scaling factor $R$ in  \eqref{eq54} is introduced, instead of a single term $r$ determined generally by \cite{yang2017dynamically}
\begin{equation}
\label{eq72}
\dot{r}=\gamma \Gamma_{r}\|\bm{\Psi}\|^{2}r,~\text{any}~\Gamma_{r}>1/(2J_{\rm m})
\end{equation}
to construct the scaled estimation error $\bm{z}$. This modification offers two prominent advantages for control design and analysis: (i) it eliminates the need for the minimum eigenvalue of $\bm{J}$ in designing the scaling factor dynamics \eqref{eq56}, as discussed in \cite{wen2018dynamic,xia2020immersion,shao2021data}; (ii) only constant gains $k_{p}$ and $k_{f}$ are chosen for the target dynamics \eqref{eq10} and the state filter \eqref{eq23}, respectively, rather than dynamic ones involving $r$ that sustains unlimited growth under perturbed and noisy conditions due to the lack of damping (see \eqref{eq72}), which avoids causing high-gain control and hence undesirable transient behavior of the closed-loop system. The latter implies that a scaling factor is not necessarily required to deal with the perturbation resulting from the regressor reconfiguration; in fact, it is introduced just for analysis and would not be used in controller implementation, thus significantly reducing the algorithm complexity.

3) Concerning the DREM-based gradient descent estimator $\dot{\hat{\bm{\theta}}}=-\Gamma\Delta(\Delta\hat{\bm{\theta}}-\bm{Y})$ derived based upon the LRE \eqref{eq34}, it has been claimed in \cite{aranovskiy2016performance,ortega2020new} that non-square integrability of the scalar regressor $\Delta$ is required to ensure asymptotic parameter convergence, which further becomes exponential by imposing the PE condition. However, both the conditions $\Delta\notin\mathcal{L}_{2}$ and $\Delta\in\mbox{PE}$ can hardly be fulfilled in practice. Actually, $\bm{W}_{a}^{\top}$ in many cases has only a weak IE, so does $\Delta$, hence the gradient estimator usually fails to achieve consistent parameter convergence. For the cases satisfying IE, a determinant detection and updating freeze mechanism, recently used in the composite learning control methods \cite{guo2020composite,dong2019composite,shao2021data}, can be introduced to ensure non-degradation of $\Delta$ after the end of excitation. But, if $\Delta$ is determined in such a manner, the estimator will lose alertness to parameter variations. To circumvent the above problems, an LTV filter \eqref{eq37} borrowed from \cite{yi2021almost} is applied to extend the scalar LREs in \eqref{eq34} to \eqref{eq40}, yielding a new scalar regressor $\Delta_{\tt N}$ that satisfies PE when $\bm{W}_{a}^{\top}$ is only of IE, as dictated by Lemma \ref{Lemma3}. Thus, the excitation requirement for exponential parameter convergence is significantly relaxed without resorting to a freeze operation. In addition, due to exponential forgetting design in  \eqref{eq31} and \eqref{eq32}, $\bm{Y}$ and $\Delta$ discount the obsolete information about the unknown parameters, in favour of new information that is conveyed by recent data. Benefiting from this property, the learning law $-\gamma\lambda\bm{\epsilon}$ in \eqref{eq46} constructed by the extended LRE \eqref{eq40} has the potential to achieve on-line identification of the time-varying inertia parameters, under a sufficient excitation. We underscore that the equality \eqref{eq39} only holds when $\bm{\theta}$ is constant, and that $\Xi$ accumulates historical data via forward integration. As a consequence, for a time-varying $\bm{\theta}$, there is a perturbation term appearing in \eqref{eq40}, leading to a parameter estimation error. Periodic re-initialization of the LTV filter \eqref{eq37} may be effective to improve the estimation accuracy, and the re-initialization rule is left for future research.


4) At this point, we shall show that the adaptive law \eqref{eq46} can be generalized to achieve finite/fixed-time parameter convergence. To start with, a continuous vector function of the following form is introduced  
\begin{equation}
\label{eq73}
\lceil\bm{x}\rfloor^{\iota}\doteq\|\bm{x}\|^{\iota}\textbf{Sign}_{\rm n}(\bm{x})
\end{equation}
where $\iota>0$ and $\textbf{Sign}_{\rm n}(\bm{x})$ is the so-called norm-normalized sign function \cite{li2021simultaneous}
\begin{equation}
\label{eq74}
\textbf{Sign}_{\rm n}(\bm{x})\doteq\left\lbrace
\begin{split}
&\dfrac{\bm{x}}{\|\bm{x}\|}, &~~\text{if}~~\bm{x}\neq \bm{0} \\
&\bm{0},                     &~~\text{if}~~\bm{x}=\bm{0}
\end{split}\right.
\end{equation}
Then, the adaptive law \eqref{eq46} is modified as
\begin{equation}
\label{eq75}
\dot{\hat{\bm{\theta}}}=-\gamma[\dot{\bar{\bm{\mu}}}-(\bm{\Phi}+\bm{\Psi})^{\top}\bar{\bm{y}}]-\gamma\left(\lambda\bm{\epsilon}+\bm{\Theta}\right) 
\end{equation}
with $\bm{\Theta}$ bing a power term defined by
\begin{equation}
\label{eq76}
\bm{\Theta}=\lambda_{1}\lceil\bm{\epsilon}\rfloor^{\iota_{1}}+\lambda_{2}\lceil\bm{\epsilon}\rfloor^{\iota_{2}}
\end{equation}
where $\lambda_{1}$,\,$\lambda_{2}>0$ are adaption gains, $\iota_{1}\in(0,1)$, and $\iota_{2}>1$. This results in slight modifications of the estimation error dynamics \eqref{eq52} and the scaled error dynamics \eqref{eq57} as follows: 
\begin{equation}
\label{eq77}
\dot{\tilde{\bm{\theta}}}=-\gamma(\bm{\Phi}+\bm{\Psi})^{\top}\bm{J}^{-1}\bm{\Phi}\tilde{\bm{\theta}}-\gamma\left(\lambda\bm{\epsilon}+\bm{\Theta}\right)
\end{equation}
\begin{equation}
\label{eq78}
\dot{\bm{z}}=-\gamma(\bm{\Phi}+\bm{\Psi})^{\top}\bm{J}^{-1}\bm{\Phi}\bm{z}-\gamma\lambda\Delta_{\tt N}\bm{z}-\gamma\dfrac{\bm{\Theta}}{R}-\dfrac{\gamma}{2J_{\rm m}}\|\bm{\Psi}\|^{2}\bm{z}
\end{equation}
As per the definitions of $\textbf{Sig}_{\rm n}^{\iota}(\cdot)$ in \eqref{eq73} and $\bm{\Theta}$ in  \eqref{eq76}, it is easy to check that $\bm{z}^{\top}\bm{\Theta}/R\geq0$ always holds for all $t\geq0$. Thus, the above modifications will not affect the results presented in Theorem \ref{Theorem1}. As for all $t\in[T_{s},\infty)$, it has $\Delta_{\tt N}(t)>\hbar$, whereby we further get
\begin{equation*}
\begin{aligned}
&\bm{z}^{\top}(\bm{\Theta}/R)= \\
&\left\lbrace
\begin{split}
&\lambda_{1}\Delta_{\tt N}^{\iota_{1}}R^{\iota_{1}-1}\|\bm{z}\|^{\iota_{1}+1}+\lambda_{2}\Delta_{\tt N}^{\iota_{2}}R^{\iota_{2}-1}\|\bm{z}\|^{\iota_{2}+1}, &\text{if}~~\tilde{\bm{\theta}}\neq\bm{0} \\
&0, &\text{if}~~\tilde{\bm{\theta}}=\bm{0}
\end{split}\right.
\end{aligned}
\end{equation*}
Consider again the Lyapunov-like function $V_{z}$ in \eqref{eq58}. Then, using the above equation in $\dot{V}_{z}$ yields
\begin{align}
\dot{V}_{z}&\leq-\lambda\hbar\|\bm{z}\|^{2}-\lambda_{1}\hbar^{\iota_{1}}R^{\iota_{1}-1}\|\bm{z}\|^{\iota_{1}+1}-\lambda_{2}\hbar^{\iota_{2}}R ^{\iota_{2}-1}\|\bm{z}\|^{\iota_{2}+1} \nonumber \\
&\leq-2\lambda\gamma\hbar V_{z}-c_{1}V_{z}^{\frac{\iota_{1}+1}{2}}-c_{2}V_{z}^{\frac{\iota_{2}+1}{2}} \label{eq80}
\end{align}
for all $t\geq T_{s}$, where $c_{1}\doteq(2\gamma)^{\frac{\iota_{1}+1}{2}}\lambda_{1}\hbar^{\iota_{1}} R_{\rm m}^{\iota_{1}-1}$ and $c_{2}\doteq(2\gamma)^{\frac{\iota_{2}+1}{2}}\lambda_{2}\hbar^{\iota_{2}}R_{\rm m}^{\iota_{2}-1}$ are defined for brevity, and the constant $R_{\rm m}>0$ denotes the minimum value of $R$ on $t\in[T_{s},\infty)$. From  \eqref{eq80} and the finite/fixed-time stability theorem \cite{yu2005continuous,zuo2017fixed}, it can be concluded that the equilibrium point $\bm{z}=\bm{0}$ of the scaled estimation error dynamics \eqref{eq78} is fixed-time (respectively, finite-time) stable on $t\in[T_{s},\infty)$, if  $\lambda_{1},\,\lambda_{2}\neq0$ (respectively, $\lambda_{1}\neq0$ and $\lambda_{2}=0$). As the scaling factor $R>0$, $\tilde{\bm{\theta}}$ also converges to zero in finite/fixed time, and the settling time $T_{f}$ for finite and fixed-time parameter convergence can be estimated, respectively, as
\begin{equation}
\label{eq81}
T_{f}\leq \dfrac{1}{\lambda\gamma\hbar(1-\iota_{1})}\ln\dfrac{2\lambda\gamma\hbar V_{z}^\frac{1-\iota_{1}}{2}(T_{s})+c_{1}}{c_{1}}
\end{equation}
\begin{equation}
\label{eq82}
T_{f}\leq T_{s}+\dfrac{2}{c_{1}(1-\iota_{1})}+\dfrac{2}{c_{2}(\iota_{2}-1)}
\end{equation}

Apart from the finite/fixed-time convergence property, an interesting phenomenon called ``synchronized convergence'' is also observed for $\tilde{\bm{\theta}}$. By synchronized convergence we mean that all the elements of $\tilde{\bm{\theta}}$ converge to zero \textit{almost} at the same time. A concise analysis is provided below to illustrate this new phenomenon. Consider that $\lim_{t\to\infty}\bm{J}^{-1}\bm{\Phi}(t)\tilde{\bm{\theta}}(t)=\bm{0}$ (see \eqref{eq66}) and the convergence can be made arbitrarily fast by tuning the learning gain $\gamma$, the first term on the RHS of  \eqref{eq77} can be guaranteed to decay much faster than $\tilde{\bm{\theta}}$. As such, we ignore this term in \eqref{eq77} showing that
\begin{equation}
\label{eq83}
\dot{\tilde{\bm{\theta}}}=-\gamma\left(\lambda\bm{\epsilon}+\bm{\Theta}\right)
\end{equation}
From  \eqref{eq76} and \eqref{eq83}, it is easy to verify that, for any $\tilde{\theta}_{i}, \tilde{\theta}_{j}\neq0$, $i\neq j$, $\frac{\rm d}{{\rm d}t}(\tilde{\theta}_{i}/\tilde{\theta}_{j})=0$ always holds, from which we claim that any nonzero $\tilde{\theta}_{i}$ and $\tilde{\theta}_{j}$, $i,j\in\{1,2,...,6\}$, $i\neq j$ are proportionable to each other. Thus, $\tilde{\theta}_{i}$ and $\tilde{\theta}_{j}$ converge to zero \textit{almost} at the same time, indicating the synchronized convergence. We should emphasize that the above analysis is established by ignoring the term $-\gamma(\bm{\Phi}+\bm{\Psi})^{\top}\bm{J}^{-1}\bm{\Phi}\tilde{\bm{\theta}}$ in  \eqref{eq77}, which makes the formula \eqref{eq83} approximately hold. This is the reason why we claim that all the elements of $\tilde{\bm{\theta}}$ converge to zero \textit{almost} at the same time. Although the new property lacks theoretical rigor to some extent, the parameter convergence rates across all components can be tuned in a well-balanced way and, therefore, is preferable for on-line identification.

\section{Numerical Simulations} \label{secV}

In this section, numerical simulations are carried out to show the effectiveness and key features of the composite I\&I adaptive control scheme developed in this paper. Consider the rigid-body attitude dynamics described by  \eqref{eq1} and \eqref{eq2}, where the inertia matrix is given by $\bm{\theta}=[20,17,15,1.4,0.9,1.2]^{\top}~\text{kg}\cdot\text{m}^2$. The reference attitude is set to \cite{yang2017dynamically}: $\bm{q}_{r}(0)=[0,0,0,1]^{\top}$ and the velocity profile $\bm{\omega}_{r}=\omega_{r}\bm{1}_{3}\,\text{rad/sec}$, where $\bm{1}_{3}=[1,1,1]^{\top}$ and $\omega_{r}$ is of the form
\begin{equation*}
\omega_{r}=0.3(1-e^{-0.01t^{2}})\cos t+te^{-0.01t^{2}}(0.08\pi+0.006\sin t)
\end{equation*}
This setting evidently generates a non-PE reference trajectory. To verify the anti-unwinding capability of the proposed adaptive control algorithm, two sets of initial body attitudes that correspond to the same physical orientation but render opposite signs of $q_{e4}(0)$ are considered in the following simulations: 
\begin{itemize}
	\item \textit{Case} 1: $\bm{q}_{v}(0)=\bm{q}_{0}$ and $\bm{q}_{4}(0)=\sqrt{1-\bm{q}_{v}^{\top}(0)\bm{q}_{v}(0)}$
	\item \textit{Case} 2: $\bm{q}_{v}(0)=-\bm{q}_{0}$ and $\bm{q}_{4}(0)=-\sqrt{1-\bm{q}_{v}^{\top}(0)\bm{q}_{v}(0)}$
\end{itemize}
with zero rate, where $\bm{q}_{0}=[0.33,-0.3,-0.62]^{\top}$. It is shown that $q_{e4}(0)>0$ for Case 1 and $q_{e4}(0)<0$ for Case 2. All the following simulations are performed using the fixed-step ODE 4 (Runge-Kutta) solver with a sample step of $0.01\,\text{sec}$.

\subsection{Nominal Performance}

In this subsection, the designed composite I\&I adaptive controller is simulated for Case 1 under a nominal (i.e., perturbation-free) scenario, in order to support the theoretical findings. The control parameters are selected as $\alpha=0.5$, $\beta=0.1$, $a=5$, $b=0.5$, $k_{\tt N}=8$, $\gamma=25$, $\lambda=0.01$, $\kappa=0.5$, and $f_{\rm m}=2$. As the reference trajectory is weakly exciting, $k_{\tt I}=1\times 10^{9}$ is introduced into  \eqref{eq35} to enhance the regressor signal strength. In addition, the initial conditions of the LTV filter \eqref{eq37} and the parameter estimator \eqref{eq46} are set as $\bm{\chi}(0)=\bm{0}$ and $\hat{\bm{\theta}}(0)+\bm{\zeta}(0)=[10,30,8,0,0,0]^{\top}$ (in fact, $\bm{\zeta}(0)=\bm{0}$), respectively. 

\begin{figure*}[hbt!]
	\centering
	\subfigure[Attitude error]{
		\includegraphics[width=6cm]{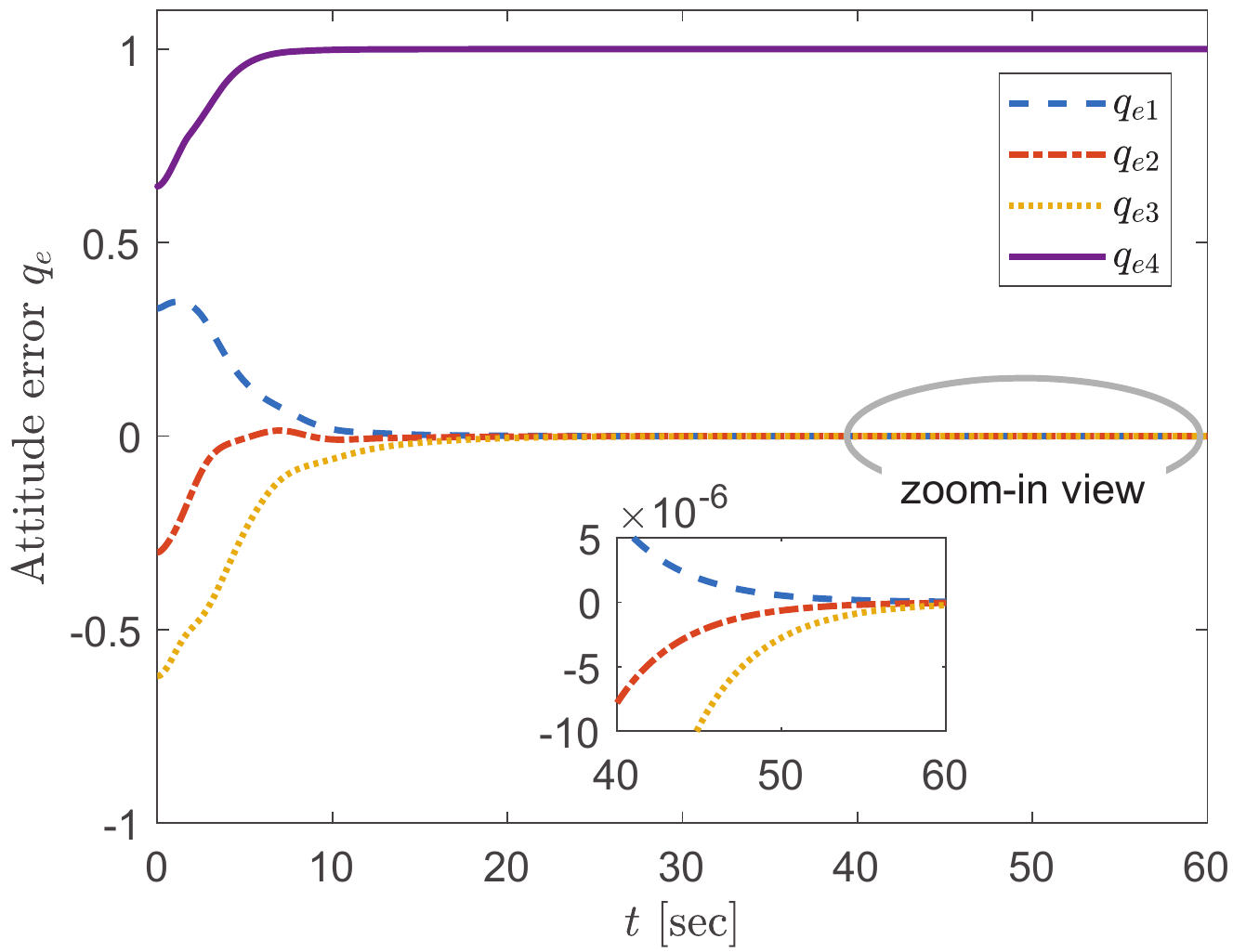}
		\label{fig1.a}}
	\subfigure[Angular velocity error]{
		\includegraphics[width=6.03cm]{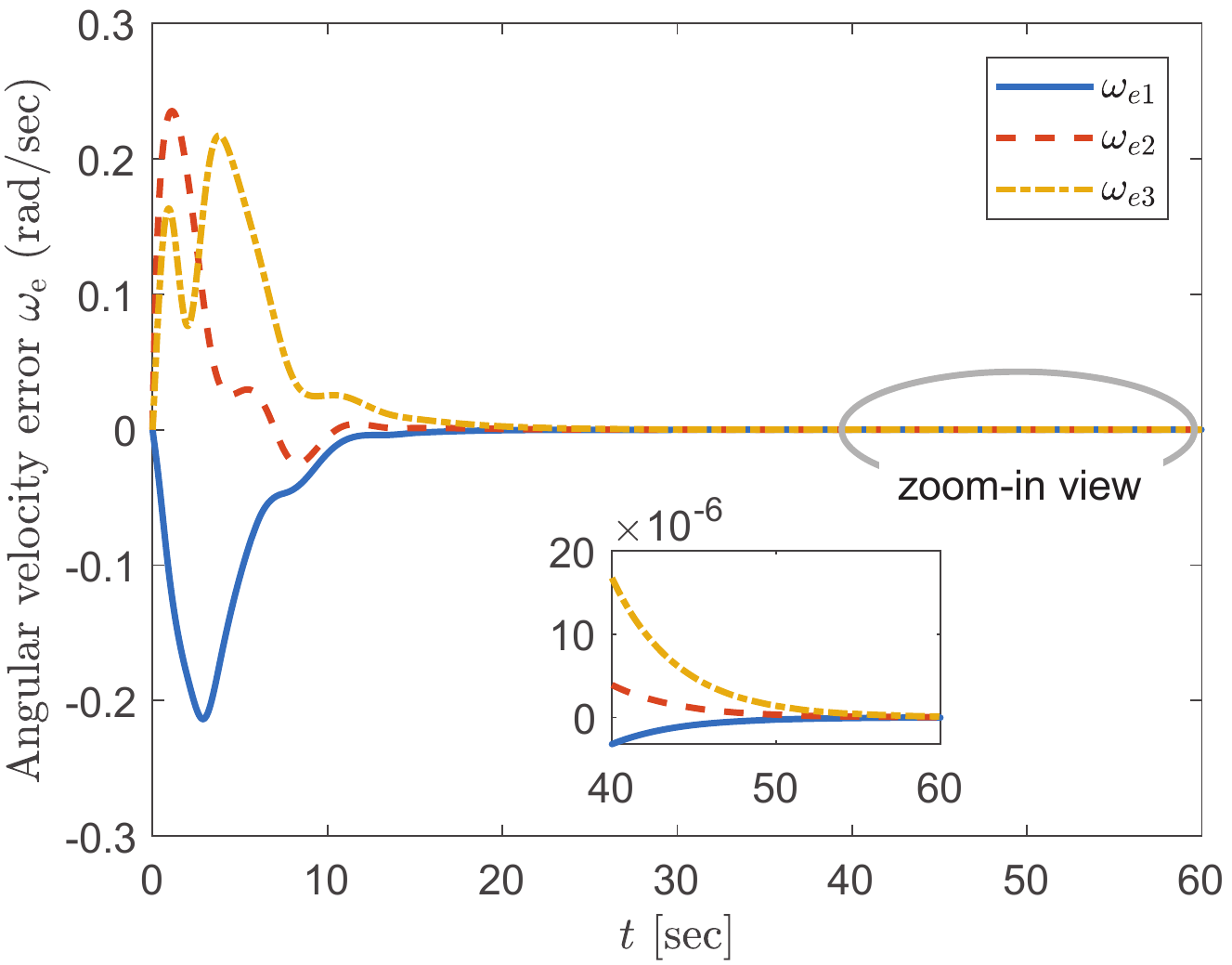}
		\label{fig1.b}}
	\subfigure[Control torque]{
		\includegraphics[width=6cm]{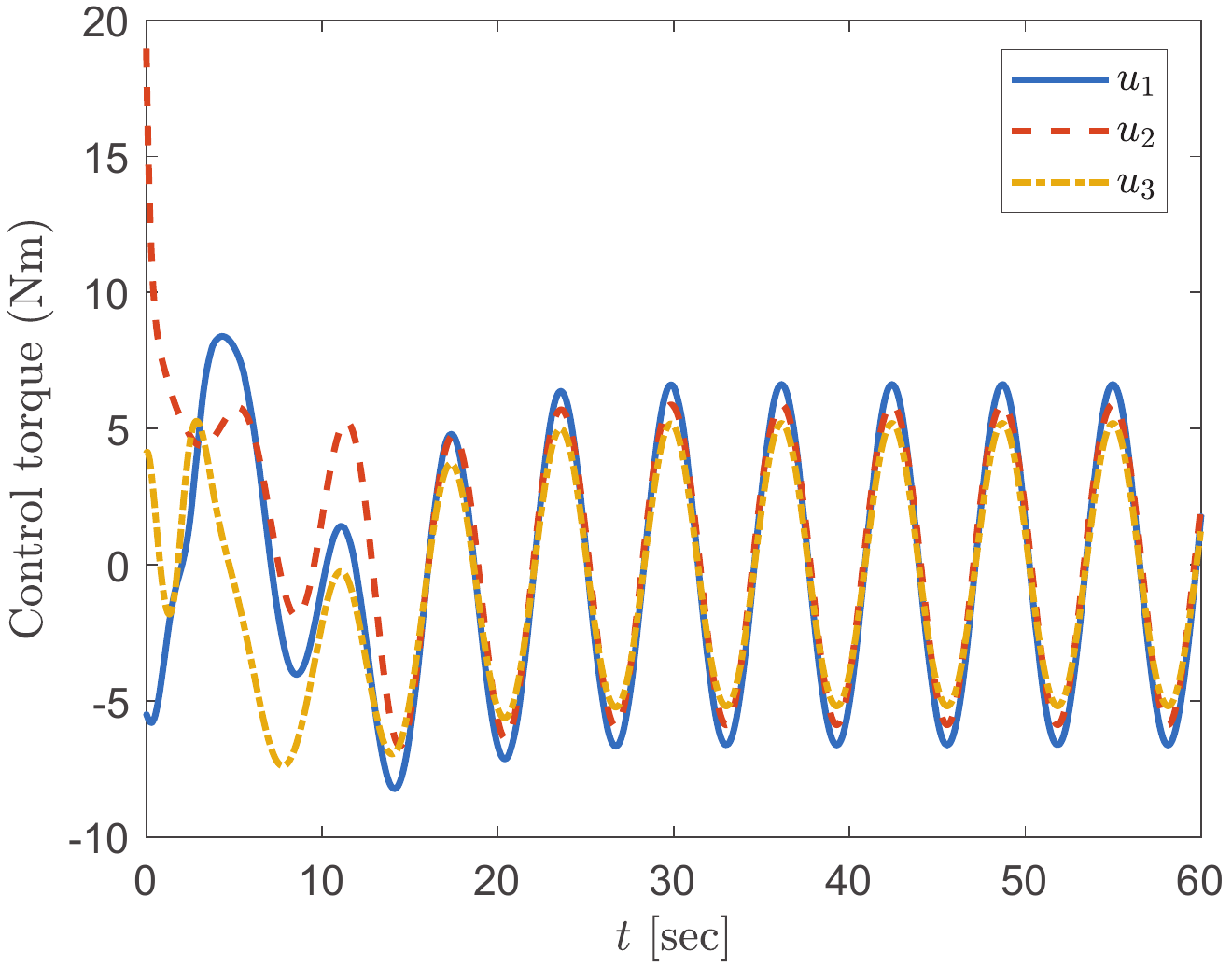}
		\label{fig1.c}}
	\subfigure[Parameter estimation]{
		\includegraphics[width=6.02cm]{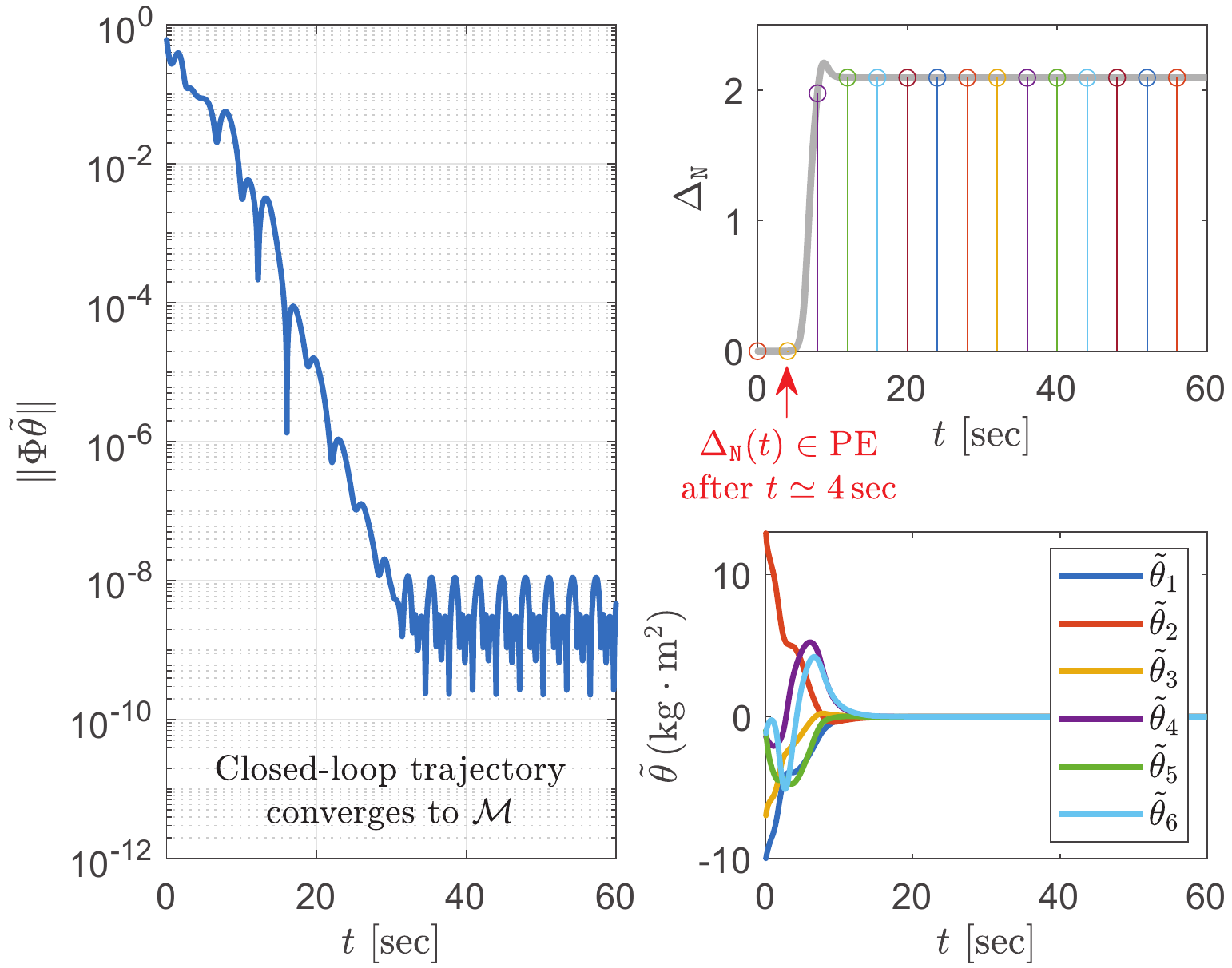}
		\label{fig1.d}}
	\caption{Control performance illustration.}
	\label{fig1}
\end{figure*}

From Figs. \ref{fig1.a} and \ref{fig1.b}, it is shown that the attitude and angular velocity tracking errors ($\bm{q}_{ev}$ and $\bm{\omega}_{e}$) converge asymptotically to zero. Note that an IE condition is satisfied during the initial phase of the mission such that $\Delta_{\tt N}(t)\in\mbox{PE}$ after $T_{s}=4\,\rm sec$, as clearly seen in Fig. \ref{fig1.d}. So, more precisely speaking, $\bm{q}_{ev}$ and $\bm{\omega}_{e}$ are exponentially convergent on $t\in[T_{s},\infty]$. The control torques commanded by the derived controller are plotted in Fig. \ref{fig1.c}, where we observe that the torque demands are smooth and remain time-varying at the steady state to ensure tracking of the assigned reference trajectory ($\bm{q}_{r}$ and $\bm{\omega}_{r}$). The time responses of $\bm{\Phi}\tilde{\bm{\theta}}$, $\Delta_{\tt N}$, and $\tilde{\bm{\theta}}$ are depicted in Fig. \ref{fig1.d}, from the left subplot of which it is clear that $\bm{\Phi}\tilde{\bm{\theta}}$ converges asymptotically to zero, indicating that the closed-loop system trajectory is indeed attracted to the invariant manifold $\mathcal{M}$. Thus, the uncertain attitude dynamics will be ultimately immersed into the target dynamics \eqref{eq10}, which contributes to the recovery of the deterministic-case closed-loop performance (no inertia uncertainties). As can be seen in the top left subplot of Fig. \ref{fig1.d}, the extended scalar regressor $\Delta_{\tt N}(t)$ turns to be strictly positive ($\implies \Delta_{\tt N}(t) \in \mbox{PE}$) after $T_{s}=4\,\rm sec$, only under an extremely weak IE condition. This result is consistent with Lemma \ref{Lemma3}, and directly helps achieve the exponential parameter and tracking error convergence on $t\in[T_{s},\infty)$, without resorting to the restrictive PE condition. The parameter convergence with good transient behaviors is observed in the bottom left subplot of Fig. \ref{fig1.d}. 

\begin{figure}[hbt!]
	\centering
	\includegraphics[width=8.3cm]{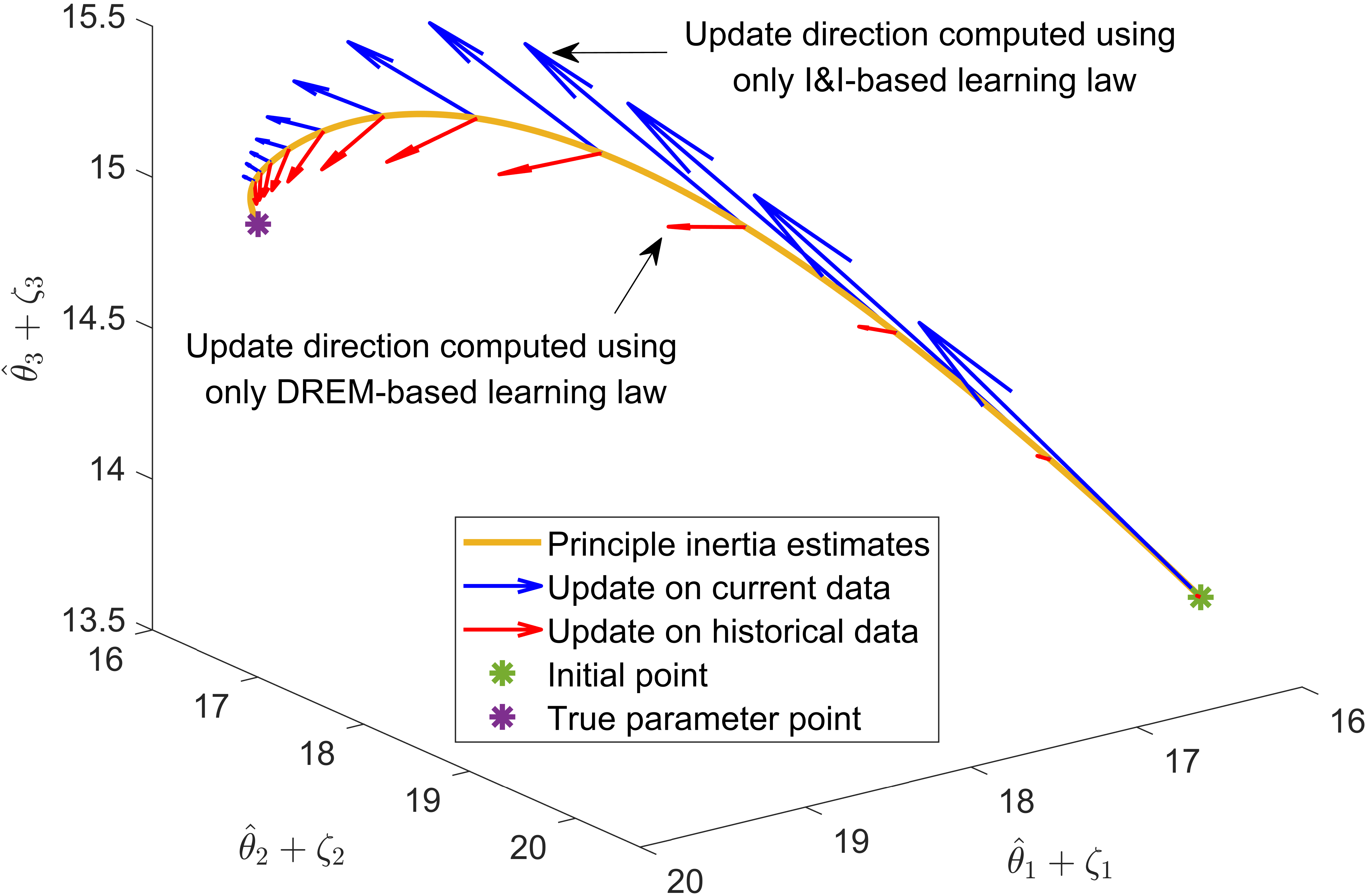}
	\caption{3-D motion trajectory of the principle inertia estimates.}
	\label{fig2}
\end{figure}

\begin{figure*}[hbt!]
	\centering
	\includegraphics[width=15cm]{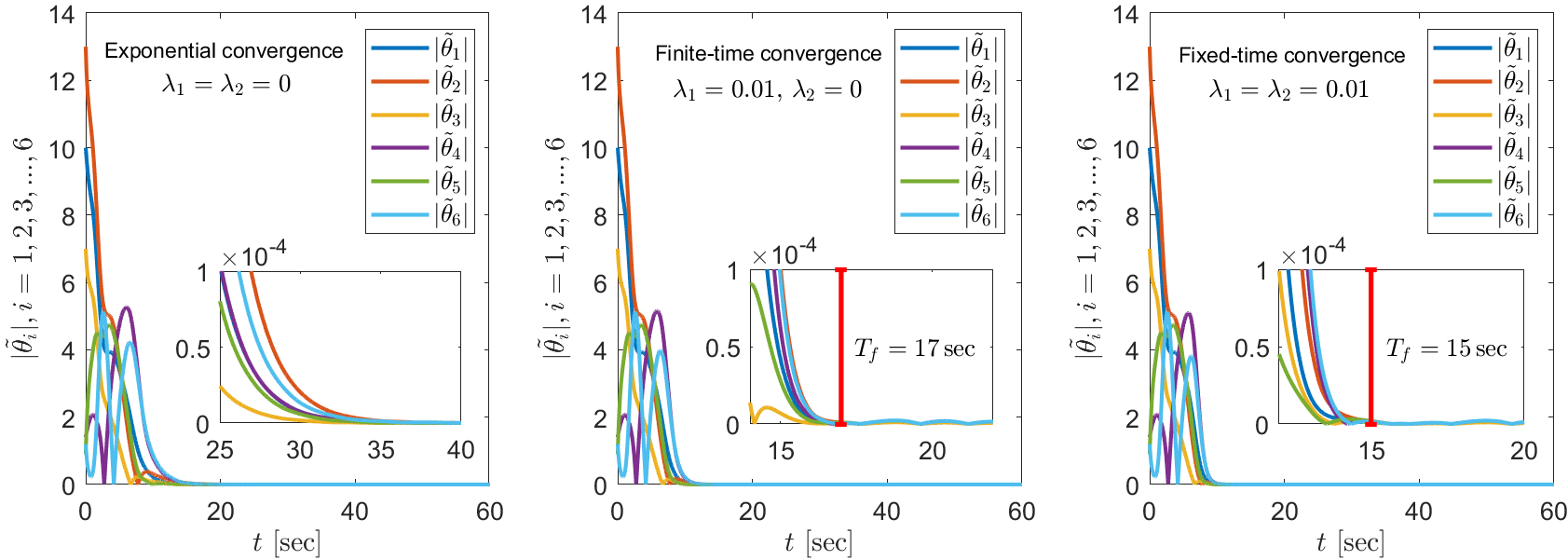}
	\caption{Time responses of $|\tilde{\theta}_{i}|$, $i=1,2,...,6$ under finite- and fixed-time adaptation extensions.}
	\label{fig3}
\end{figure*}

To examine the role played by the DREM-based learning law $-\gamma\lambda\bm{\epsilon}$ on parameter convergence, the three-dimensional (3-D) motion trajectory of the principle inertia estimates is provided in Fig. \ref{fig2}, in which the blue and red arrows show the parameter update information (directions and magnitudes) based only on the I\&I-based learning law and only on the DREM-based learning law, respectively, at regular intervals. Please notice that even though $T_{s}=4\,\rm sec$, $\Delta_{\tt N}$ is very small at $T_{s}$ and gradually increases until to $t\simeq 12\,\rm sec$. Bearing this in mind, the initial point in Fig. \ref{fig2} is taken at $t=5\,\rm sec$ for more clear illustration. By observing the arrows in Fig. \ref{fig2}, we intuitively see that the two learning laws synchronously drive the estimates of the principle inertia parameters in two linearly independently directions to their true values. This shows the importance of the DREM-based learning law in ensuring parameter convergence, in the absence of PE. To further justify the finite/fixed-time convergence property of the generalized parameter estimator \eqref{eq75}, we simulate the finite- and fixed-time adaption cases in which the gains in $\bm{\Theta}$ are set to $\lambda_{1}=0.01$, $\lambda_{2}=0$ and $\lambda_{1}=\lambda_{2}=0.01$, respectively, and the power of numbers are identically chosen as $\iota_{1}=0.85$ and $\iota_{2}=1.1$. The time responses of $|\tilde{\theta}_{i}|$, $i=1,2,...,6$ under finite- and fixed-time adaptation extensions are depicted in Fig. \ref{fig3}. The estimation results obtained from the original estimator is also provided (see the left subplot of Fig. \ref{fig3}) to serve as baseline for convergence time comparisons. As clearly seen in Fig. \ref{fig3}, the inclusion of $\bm{\Theta}$ in the adaptive law allows us to achieve \textit{synchronized} parameter convergence in a finite time, and moreover, the fixed-time adaptation extension delivers faster convergence rate than the finite-time one.    

\subsection{Comparison Results} \label{Sec.V-B}

A nominal scenario is likewise considered here, whereas the initial body attitude is changed to Case 2 to further show the efficiency of the proposed composite I\&I adaptive controller (denoted as CI\&IAC) in unwinding phenomenon avoidance. For comparison purposes, apart from the CI\&IAC, two classical adaptive controllers are also simulated:

\begin{enumerate}
	\item \textit{Filter-based Non-CE Adaptive Controller in \cite{yang2017dynamically} (denoted as NCEAC):} This controller is derived using the I\&I adaptive control method with dynamic scaling, which has been shown to offer some advantages over most existing adaptive control algorithms. The reader is referred to \cite{yang2017dynamically} for the design details of NCEAC, and the design parameters are chosen as $k_{p}=0.48$, $k_{v}=1$, $k_{r}=0.2$, $k=0.01$, $k_{H}=0.1$, $k_{1}=k_{2}=k_{3}=1$, $\Gamma=50\textbf{I}_{3}$, $v=0.5$, and $\epsilon=0.00001$. Note that although the proposed CI\&IAC is partially inspired by the NCEAC, it gives substantial improvements. As dictated by Theorem \ref{Theorem1}, the CI\&IAC preserves all the key features of the NCEAC, while achieving exponential parameter and tracking error convergence without causing unwinding.
	
	\item \textit{CE-based Adaptive Controller in \cite{costic2001quaternion} (denoted as CEAC):} The CEAC with full-state feedback not only achieves asymptotic attitude tracking, but is capable of avoiding the unwinding phenomenon by introducing a potential function $\frac{1}{2}\bm{q}_{ev}^{\top}\bm{q}_{ev}/(1-\bm{q}_{ev}^{\top}\bm{q}_{ev})$ to ensure that $q_{e4}(t)\neq0$ for all $t\geq0$. The structure of the CEAC is detailed in \cite{costic2001quaternion} with its design parameters given as $\alpha=10$, $K=10$, and $\Gamma=0.02$. 	
\end{enumerate}

It is noteworthy that the initial values of the parameter estimates for the above two controllers are chosen the same as that of the proposed CI\&IAC. In addition, to permit a fair comparison, the design parameters of the NCEAC and the CEAC are judiciously tuned by trial and error to obtain similar tracking error convergence rates as the CI\&IAC.

\begin{figure*}[hbt!]
	\centering
	\subfigure[Attitude error norm]{
		\includegraphics[width=6cm]{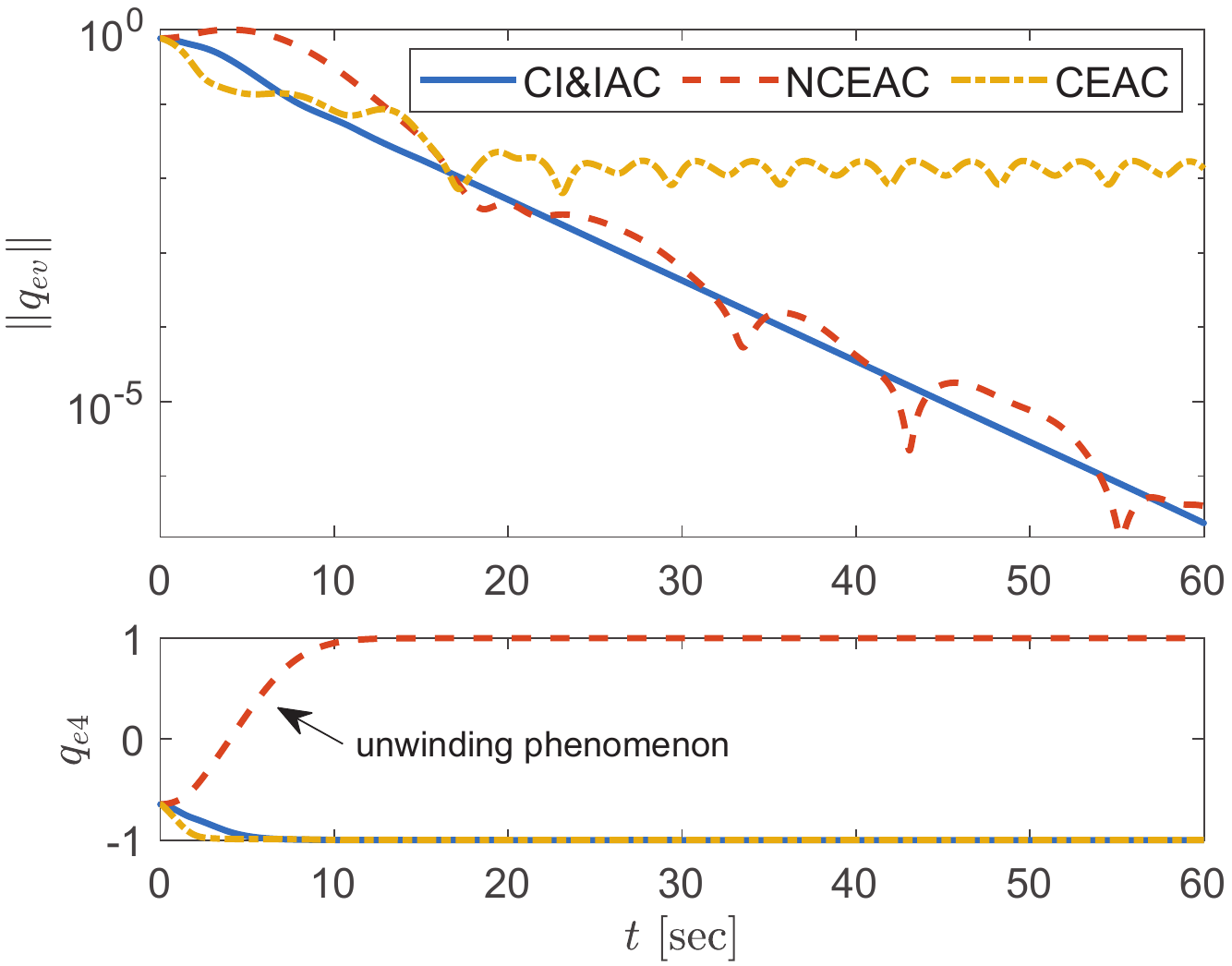}
		\label{fig4.a}}
	\subfigure[Angular velocity error norm]{
		\includegraphics[width=6cm]{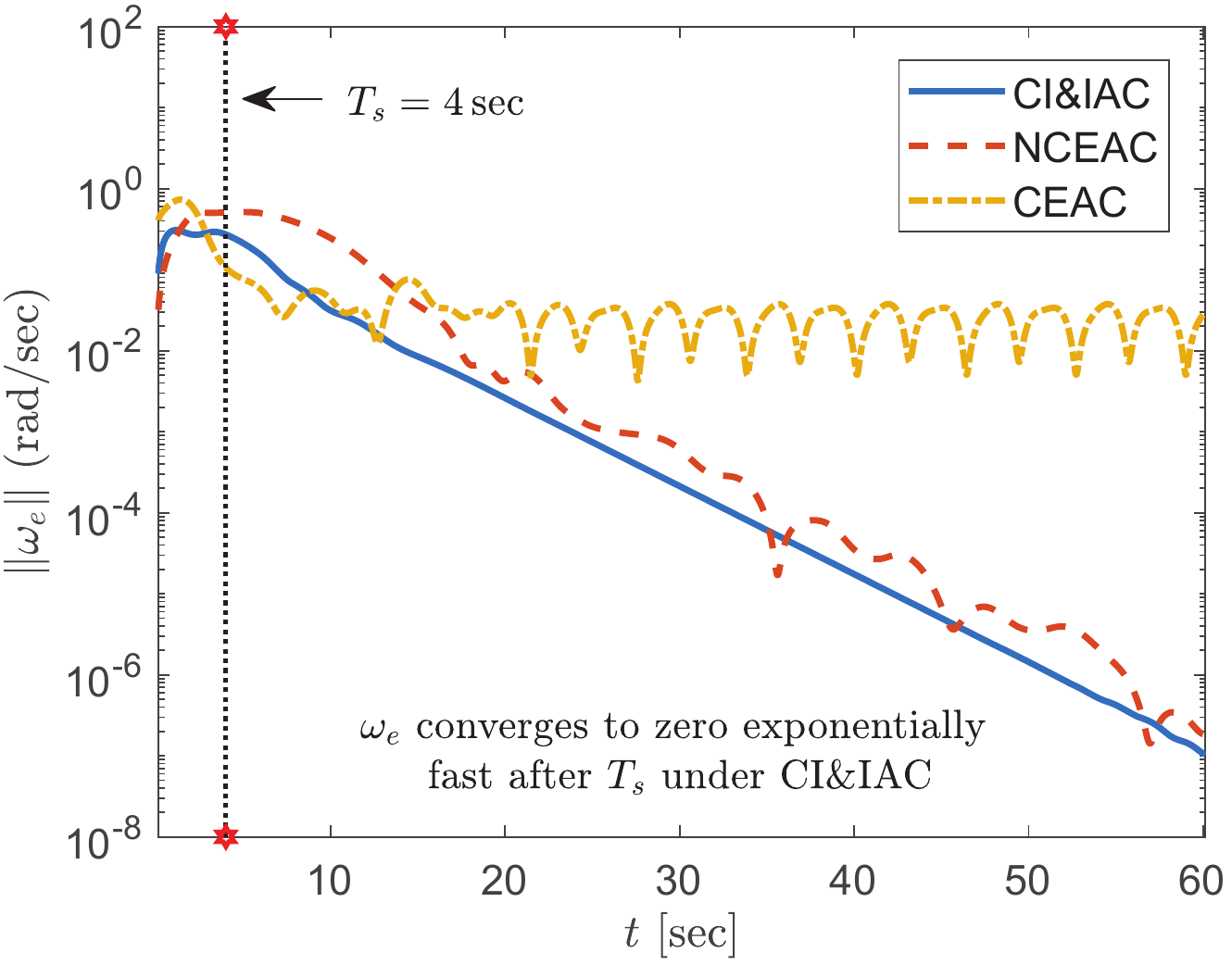}
		\label{fig4.b}}
	\subfigure[Control torque norm]{
		\includegraphics[width=6.17cm]{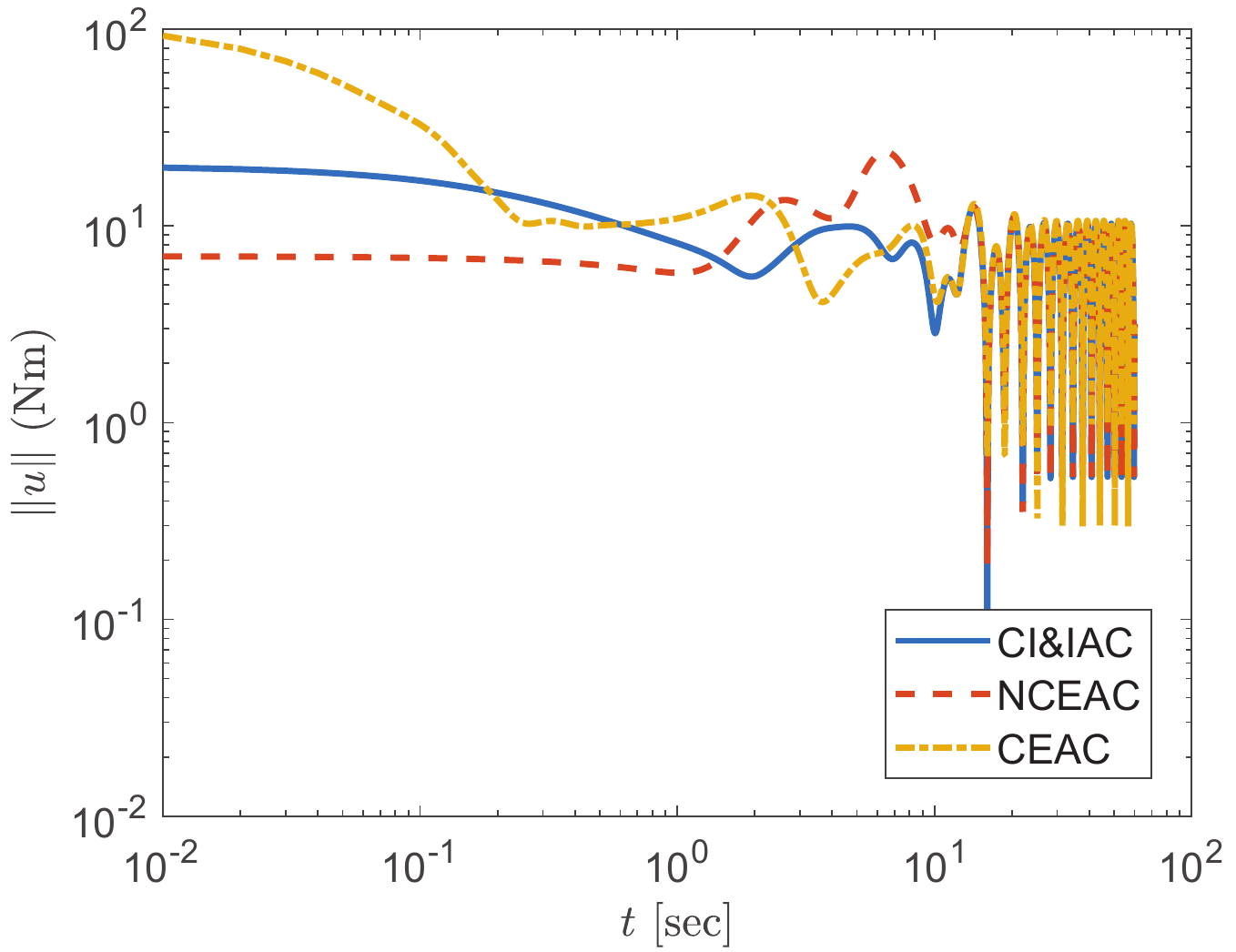}
		\label{fig4.c}}
	\subfigure[Parameter estimation error norm]{
		\includegraphics[width=6cm]{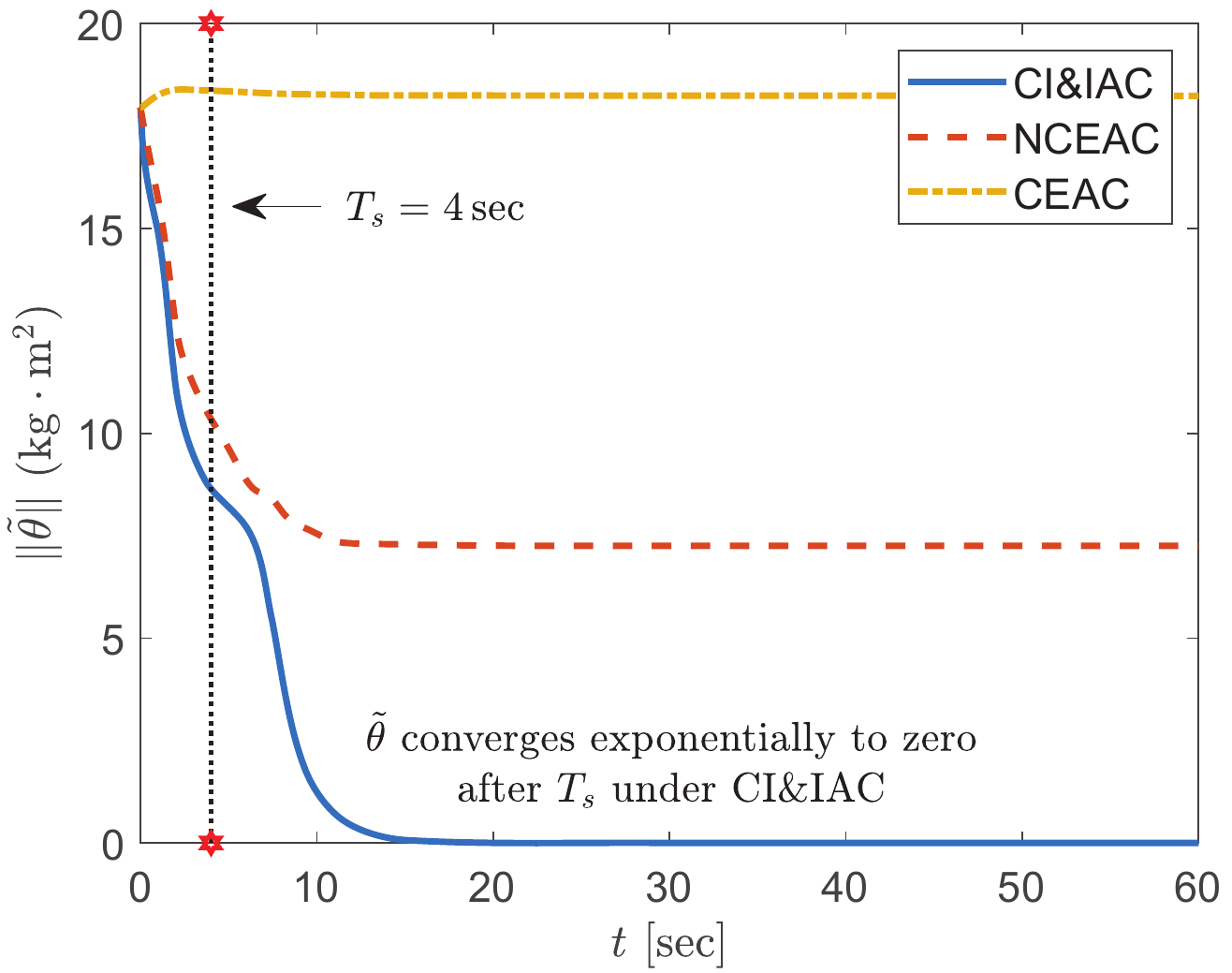}
		\label{fig4.d}}
	\caption{Control performance comparisons of different controllers under the nominal scenario.}
	\label{fig4}
\end{figure*}

As can be seen in Figs. \ref{fig4.a} and \ref{fig4.b}, all the three controllers achieve asymptotic convergence of the tracking errors $\bm{q}_{ev}$ and $\bm{\omega}_{e}$, but quantitatively speaking, the proposed CI\&IAC delivers the best transient performance, as it ensures UES of the closed-loop system after $T_{s}=4\,\rm sec$, which is not the case for the NCEAC and the CEAC. Note that $\bm{q}_{ev}$ and $\bm{\omega}_{e}$ under the CEAC exhibit very slow convergence trends, and a long simulation time is needed to see their asymptotically convergent behaviors. This is because the CE-based estimator cannot guarantee parameter convergence due to nonsatisfaction of the PE condition, which in turn renders the control performance of the CEAC to be arbitrarily poor. In contrast, the CI\&IAC and the NCEAC deviate significantly from the CE principle and effectively overcome the above deficient inherent in the CEAC, through introducing an invariant attracting manifold. Thus, they delivers satisfactory tracking performance even in the absence of PE condition. In addition, from the bottom subplot of Fig. \ref{fig4.a}, we find that both the CI\&IAC and the CEAC drive $\bm{q}_{e}$ to converge to the nearest equilibrium $[\bm{0},-1]^{\top}$, while the NCEAC fails, instead it steers $\bm{q}_{e}$ to $[\bm{0},1]^{\top}$, leading to the unwinding phenomenon (as will soon be witnessed in Fig. \ref{fig6}). The norms of the control torques commanded by the three controllers are plotted on a semilogarithmic scale as shown in Fig. \ref{fig4.c}, from which it is observed that the torque demands due to the CI\&IAC and especially the CEAC are higher than that of the NCEAC during the initial transient. This may be caused by the introduction of potential functions in the CI\&IAC and the CEAC to achieve unwinding avoidance. The comparison results in terms of parameter estimation error are depicted in Fig. \ref{fig4.d}, in which we recognize that the CI\&IAC achieves exponential parameter convergence after $T_{s}=4\,\rm sec$, without PE, while the other two controllers fails to obtain such a result, indicating that the DREM-based learning law $-\gamma\lambda\bm{\epsilon}$ is instrumental for relaxing the PE condition.     

\begin{figure*}[hbt!]
	\centering
	\subfigure[CI\&IAC]{
		\includegraphics[width=4.2cm]{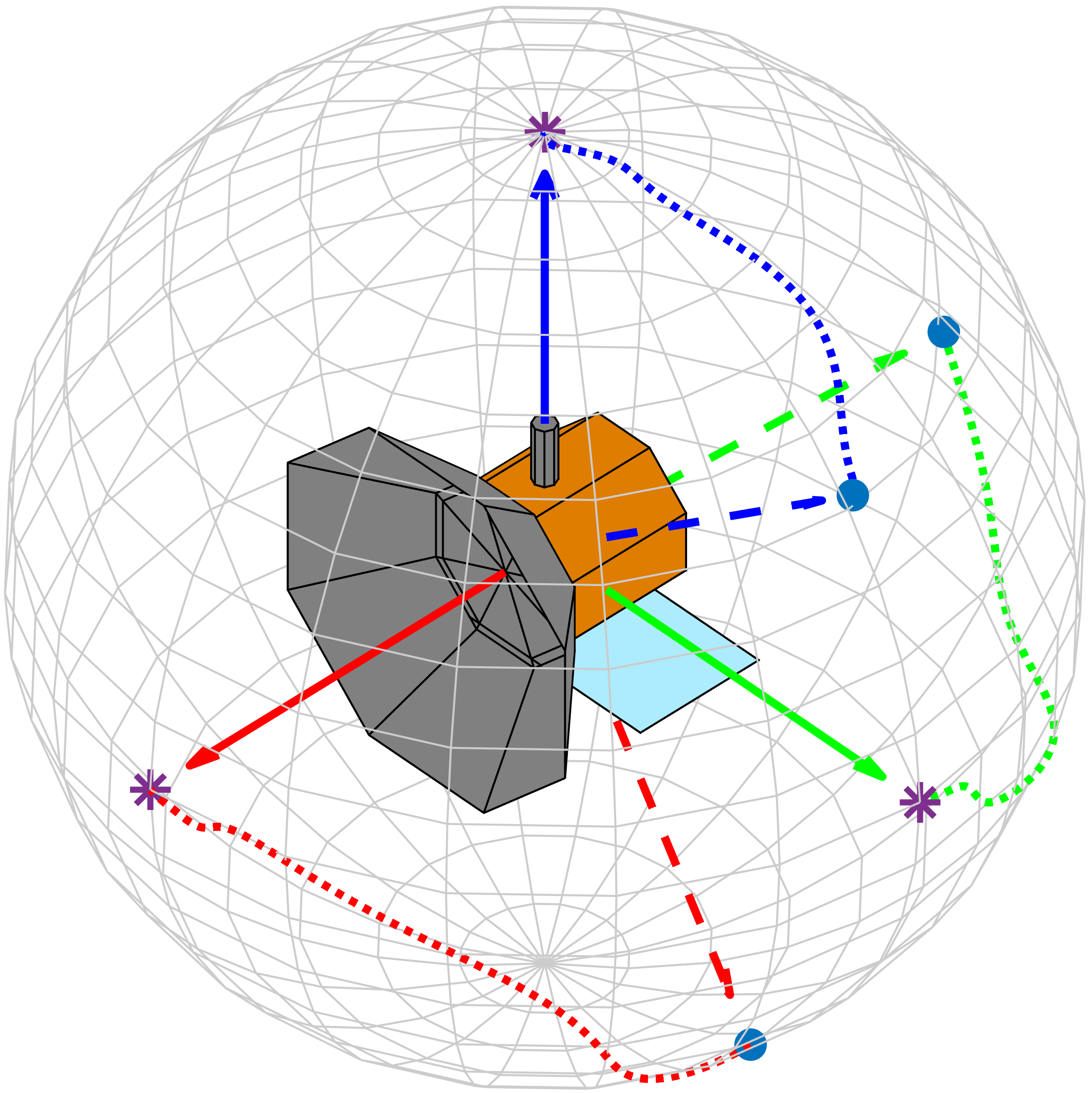}
		\label{fig5.a}}
	\subfigure[NCEAC]{
		\includegraphics[width=4.2cm]{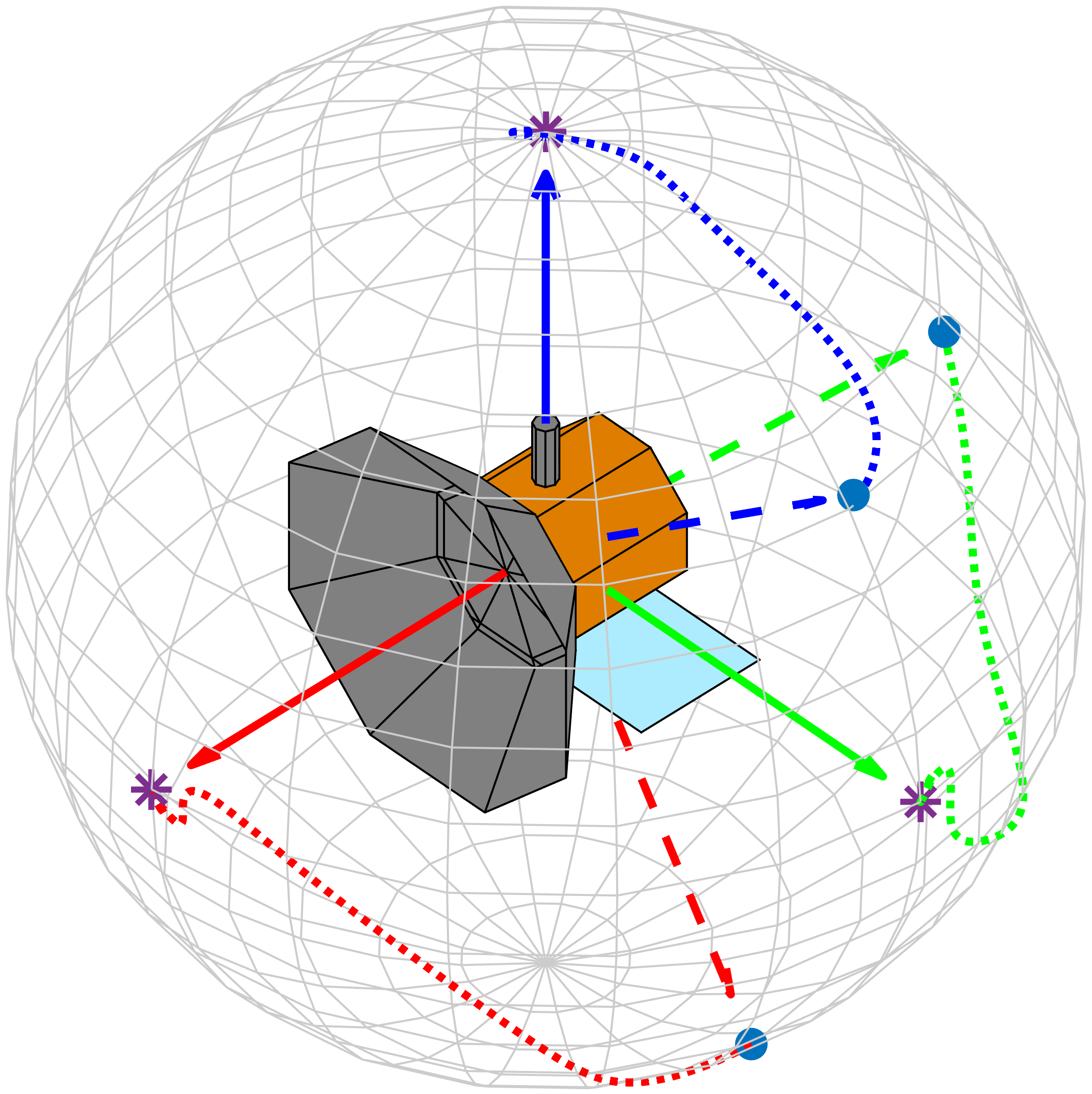}
		\label{fig5.b}}
	\subfigure[CEAC]{
		\includegraphics[width=4.2cm]{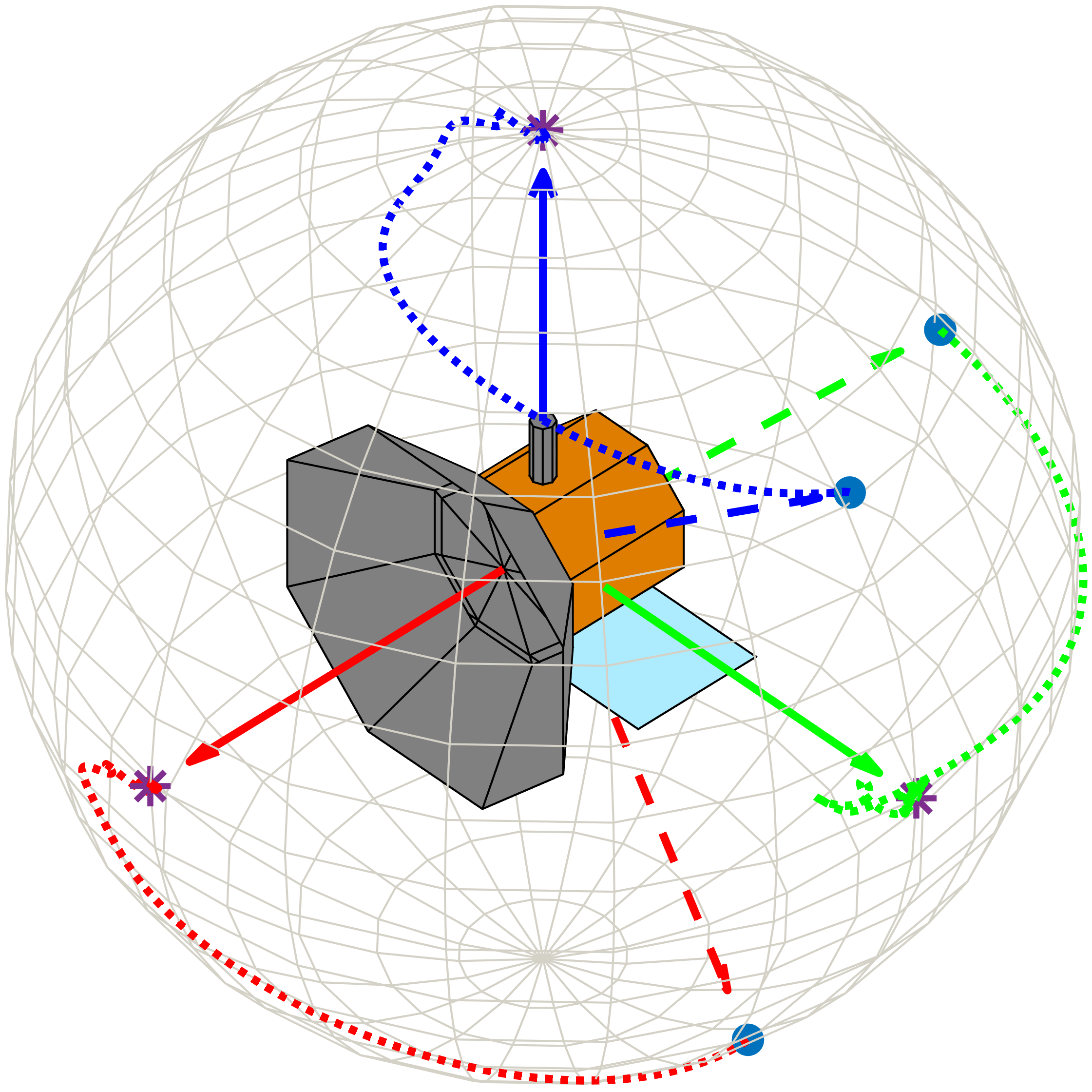}
		\label{fig5.c}}
	\caption{3-D attitude tracking trajectories observed in $\mathcal{F}_{\mathcal{R}}$ for Case 1.}
	\label{fig5}
\end{figure*}

\begin{figure*}[hbt!]
	\centering
	\subfigure[CI\&IAC]{
		\includegraphics[width=4.2cm]{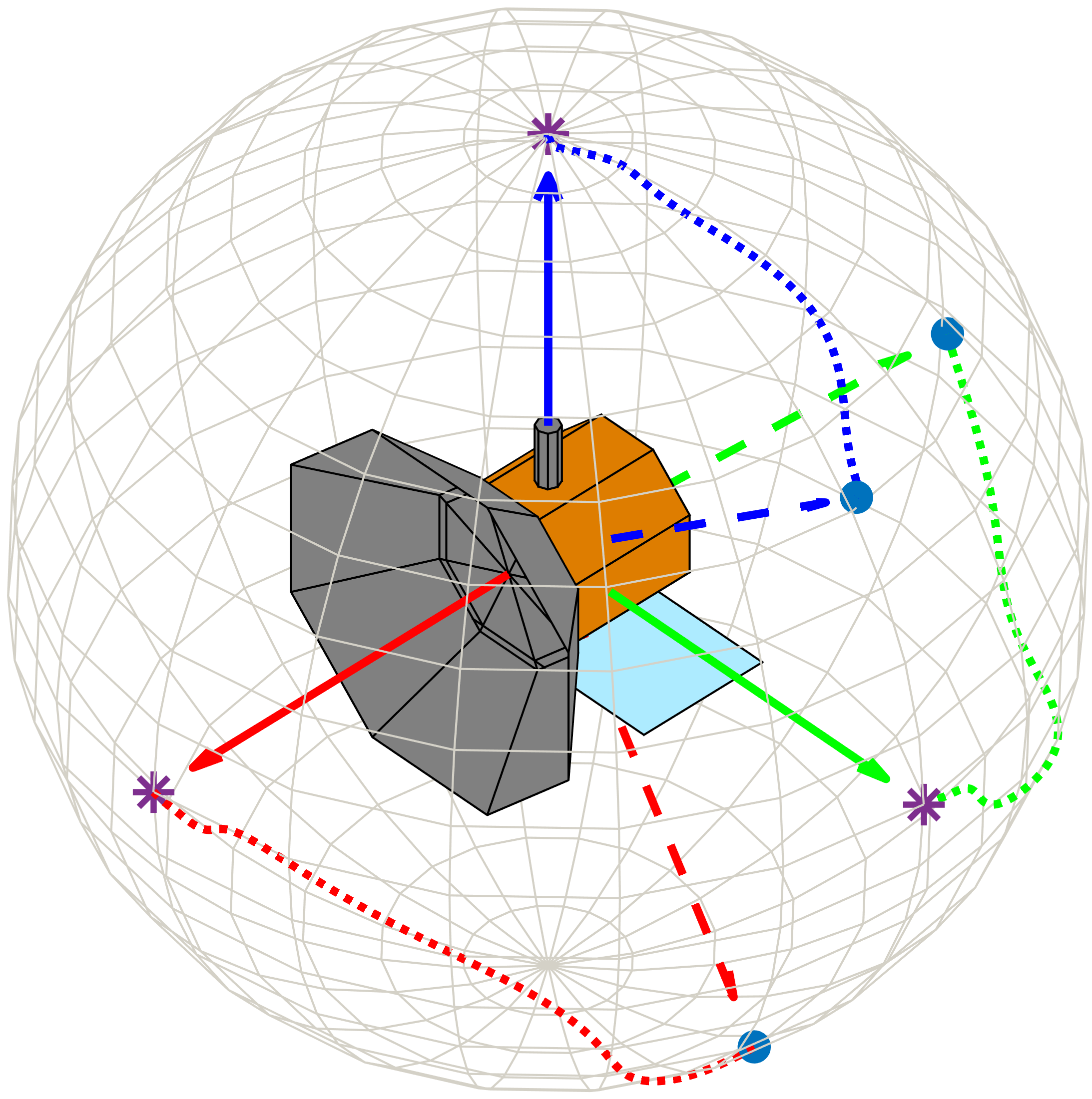}
		\label{fig6.a}}
	\subfigure[NCEAC]{
		\includegraphics[width=4.2cm]{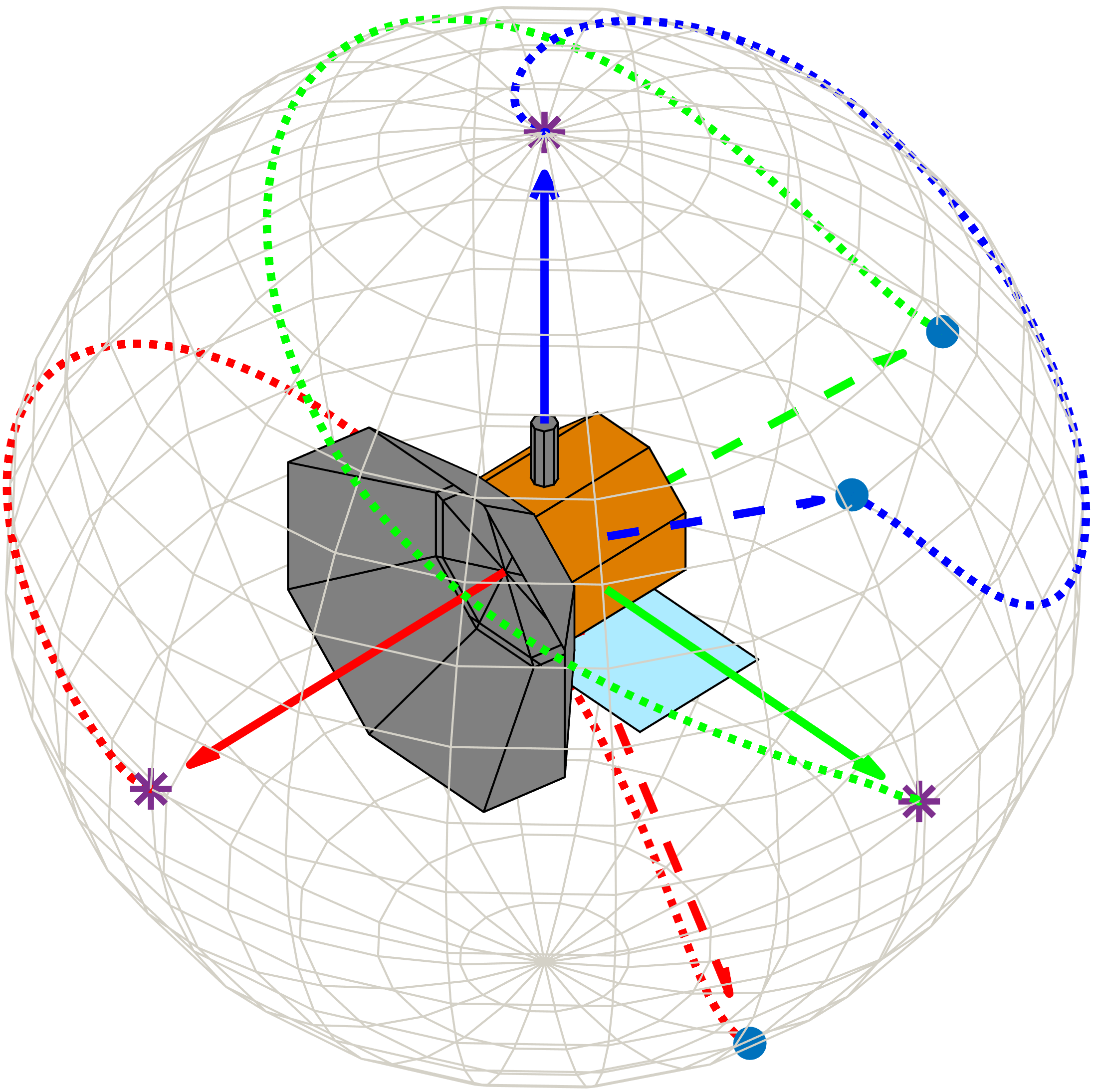}
		\label{fig6.b}}
	\subfigure[CEAC]{
		\includegraphics[width=4.2cm]{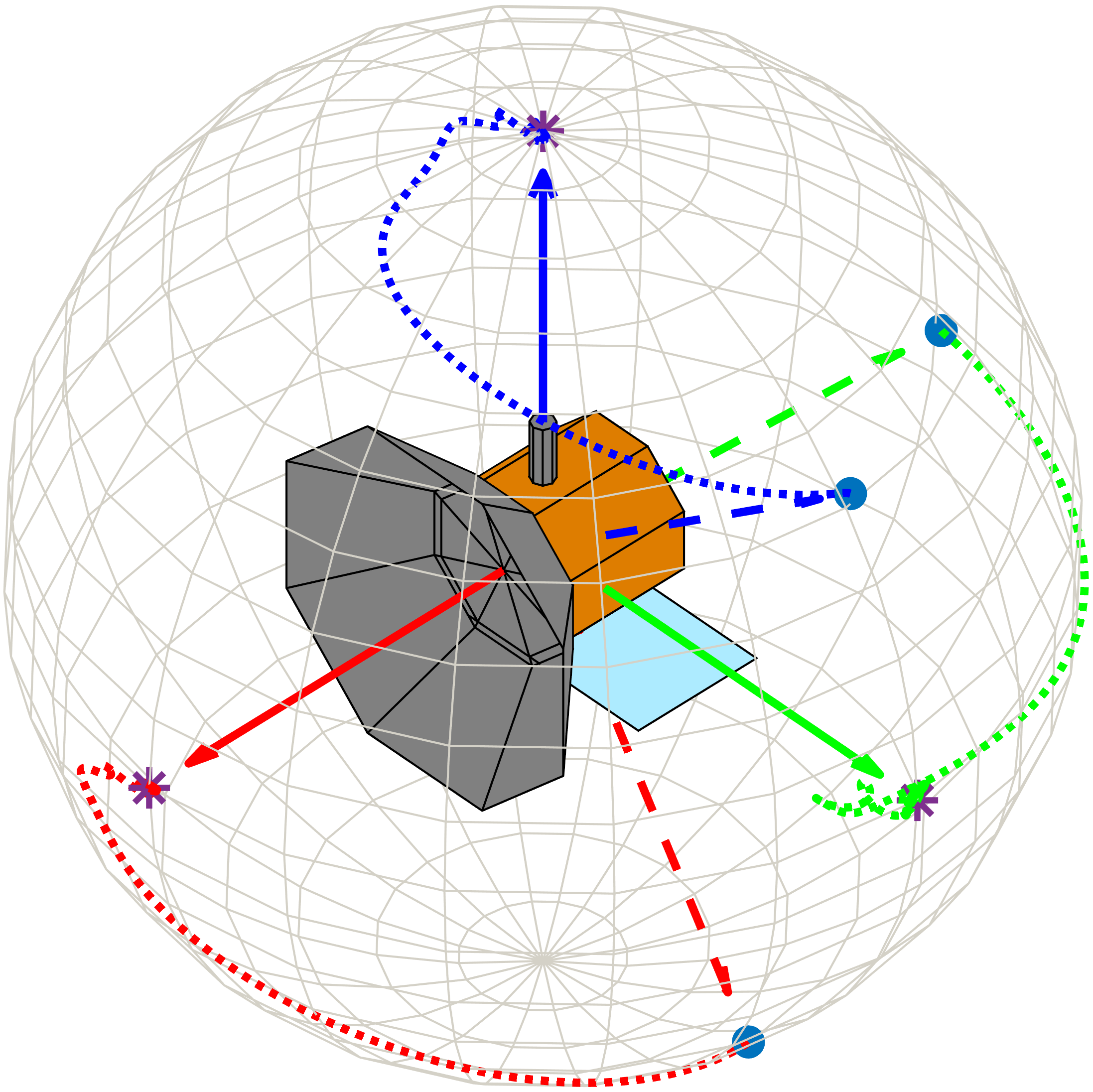}
		\label{fig6.c}}
	\caption{3-D attitude tracking trajectories observed in $\mathcal{F}_{\mathcal{R}}$ for Case 2.}
	\label{fig6}
\end{figure*}

The 3-D motion trajectories of the body frame $\mathcal{F}_{\mathcal{B}}$ w.r.t. the reference frame $\mathcal{F}_{\mathcal{R}}$ observed in $\mathcal{F}_{\mathcal{R}}$ are provided in Figs. \ref{fig5} and \ref{fig6} to better illustrate the attitude tracking processes of the three controllers under Cases 1 and 2. In Figs. \ref{fig5} and \ref{fig6}, the mutually perpendicular solid lines (red, green, and blue) denote the axes of $\mathcal{F}_{\mathcal{R}}$, while the dashed counterparts denotes the initial axes of $\mathcal{F}_{\mathcal{B}}$; moreover, the initial and desired attitude orientations are marked by ``solid dot'' and ``asterisk'', respectively. Inspecting Fig. \ref{fig5} reveals that, for Case 1 ($q_{e4}(0)>0$), all three controllers can track the prescribed reference trajectory via a rotation less than $180^{\circ}$, but intuitively, the proposed CI\&IAC renders a smoother trajectory with less fluctuation when compared with the NCEAC and the CEAC. For Case 2 ($q_{e4}(0)<0$), it is clearly shown from Fig. \ref{fig6} that the tracking trajectories from the CI\&IAC and the CEAC remain the same as in Fig. \ref{fig5}, without suffering from the unwinding phenomenon. However, the NCEAC exhibits a unnecessarily long rotation path due to unwinding.

\subsection{Robustness Validations}

At this point, we examine the robustness of the simulated controllers against external disturbances and measurement noises. The disturbance of the following form \cite{shao2021data}
\begin{equation}
\nonumber
\bm{u}_{d}=10^{-4}\times\left[
\begin{matrix}
3\cos(0.2t)+4\sin(0.06t)-10\\
-1.5\sin(0.04t)+3\cos(0.1t)+15\\
3\sin(0.2t)-8\sin(0.08t)+5
\end{matrix}\right]\text{Nm} 
\end{equation}
is introduced into the attitude dynamics \eqref{eq2}. The attitude measurement noises are modeled following the method in \cite{akella2015partial}. Towards this end, we rewrite the body attitude as $\bm{q}=[\hat{\bm{n}}^{\top}\sin(\psi/2),\cos(\psi/2)]^{\top}$, where $\hat{\bm{n}}$ and $\psi$ are known as the Euler eigenaxis and eigenangle, respectively. Within this setting, the noisy measurements of $\bm{q}$ are generated by randomly perturbing the true $\hat{\bm{n}}$ with uniform distribution in a spherical cone centered around it. The cone half-angle is set here to $0.1\,\text{deg}$. In addition, the measurement noises with mean zero and standard deviation $10^{-3}\,\text{rad/sec}$ are added to the feedback of $\bm{\omega}$. The simulation scenario in Sec. \ref{Sec.V-B} is repeated under simultaneous consideration of both external disturbances and measurement noises described above. In order to clearly illustrate the comparison results in terms of steady-state performance, the simulation duration is prolonged to $100\,\rm sec$.

\begin{figure*}[hbt!]
	\centering
	\subfigure[Attitude error norm]{
		\includegraphics[width=6cm]{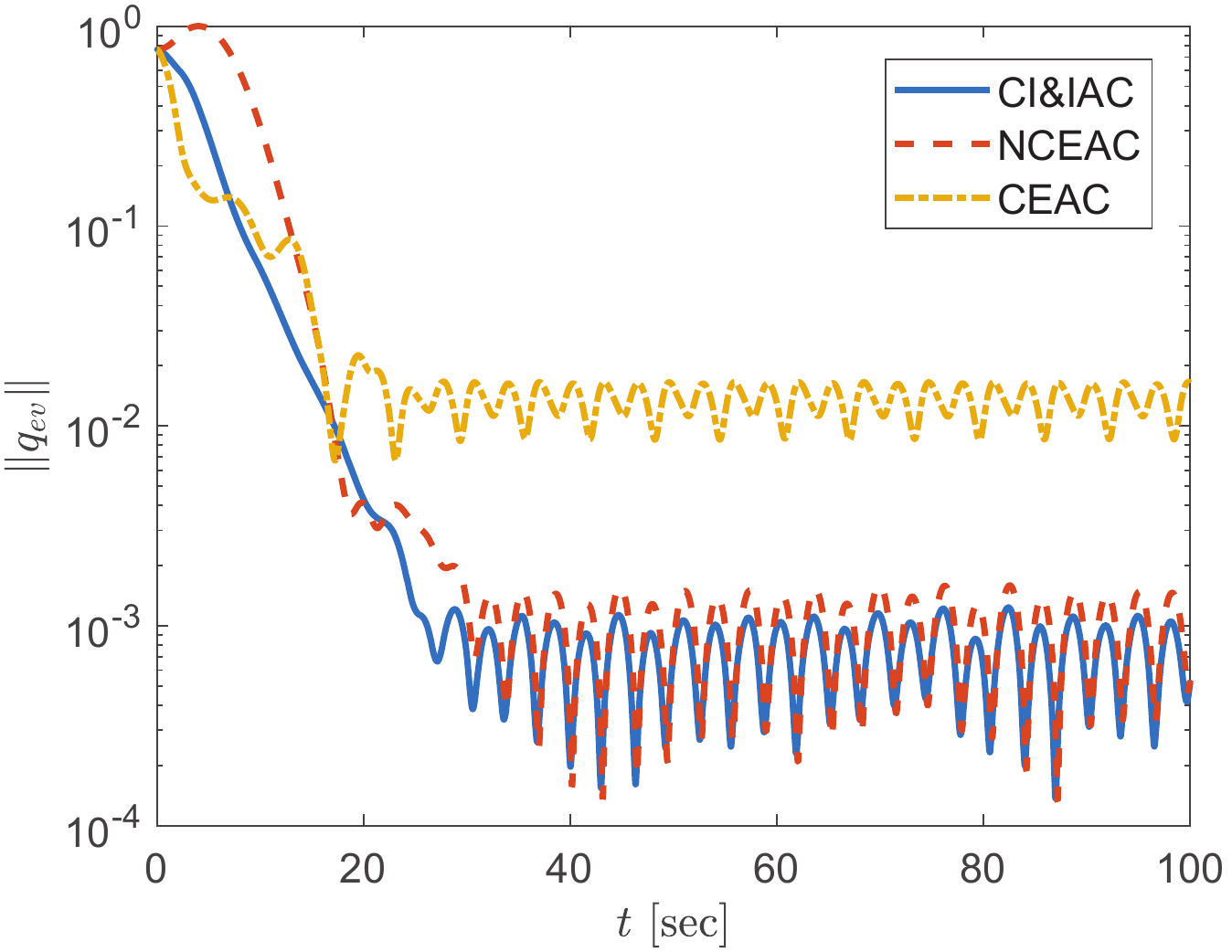}
		\label{fig7.a}}
	\subfigure[Angular velocity error norm]{
		\includegraphics[width=6cm]{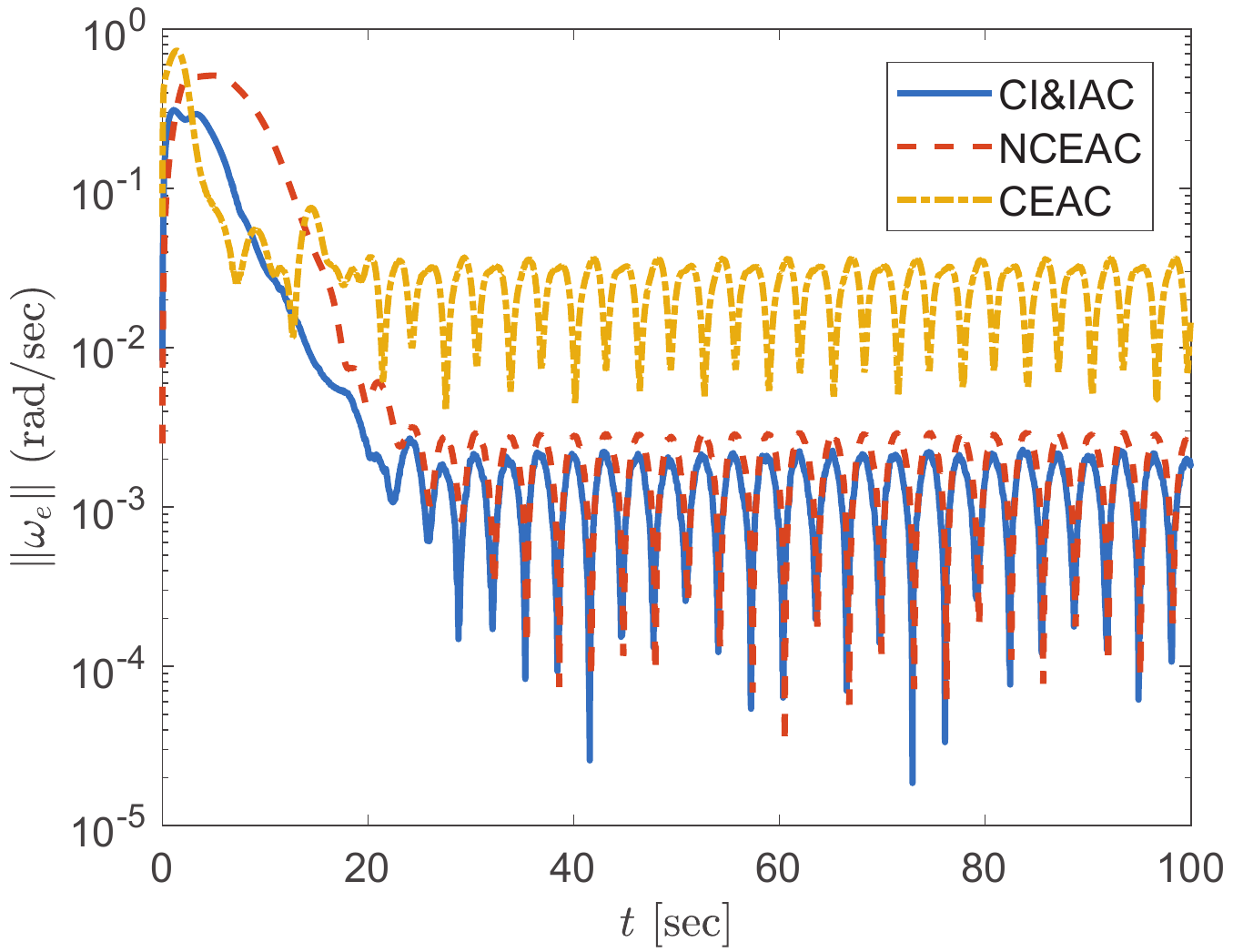}
		\label{fig7.b}}
	\subfigure[Control torque norm]{
		\includegraphics[width=6.17cm]{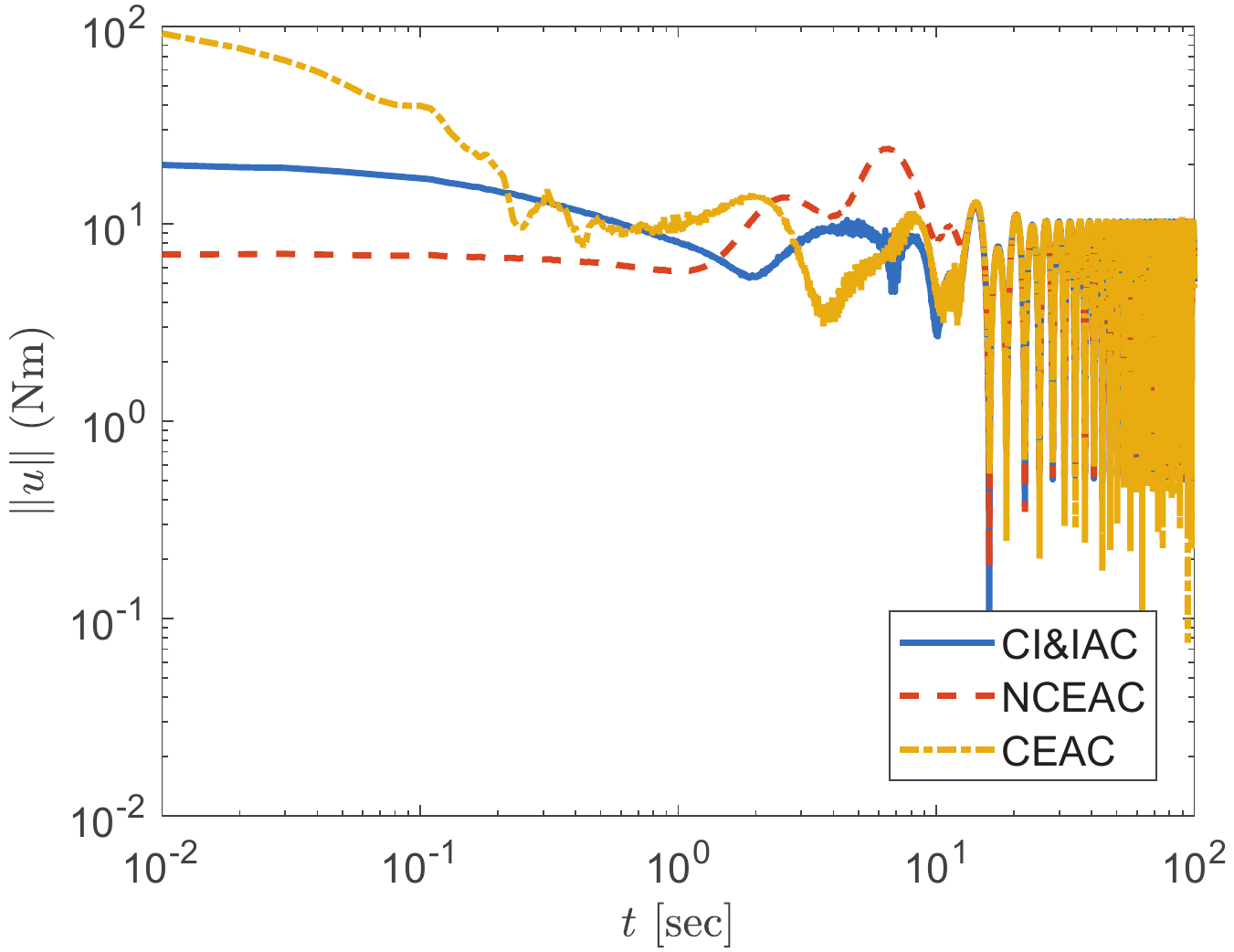}
		\label{fig7.c}}
	\subfigure[Parameter estimation error norm]{
		\includegraphics[width=6cm]{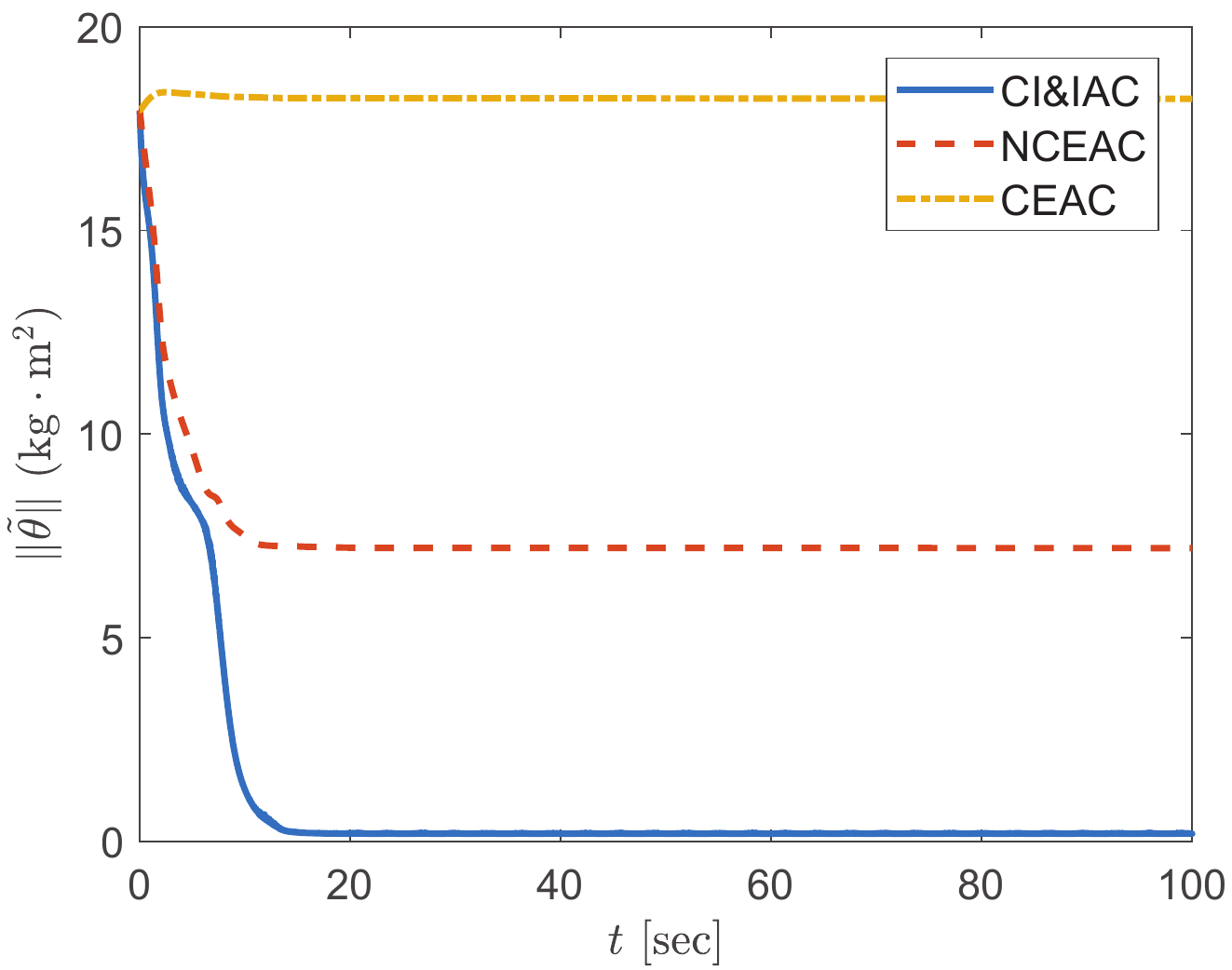}
		\label{fig7.d}}
	\caption{Control performance comparisons of different controllers under the perturbed scenario.}
	\label{fig7}
\end{figure*}

The performance comparisons of different controllers under the perturbed scenario are presented in Fig. \ref{fig7}. By comparing Figs. \ref{fig7.a} and \ref{fig7.b} with Figs. \ref{fig4.a} and \ref{fig4.b}, we find that all the controllers suffer in performance degradation, especially the steady-state accuracy, when external disturbances and measurement noises are present. The attitude and angular velocity tracking errors only converge to small residual sets around the origin, rather than zero. From a practical viewpoint, the I\&I-based adaptive controllers -- CI\&IAC and NCEAC -- still exhibit acceptable steady-state performance and outperform the CEAC. We further underscore that the proposed CI\&IAC preserves the transient-state behaviors, that is, it can steer the tracking error to converge exponentially fast to the steady-state values, and moreover, it performs slightly better than the NCEAC. The control torque norms of all three controllers are depicted in Fig. \ref{fig7.c}, in which some burrs are observed due to the noisy feedback signals. From Fig. \ref{fig7.d}, it is clear that the CI\&IAC can still ensure parameter convergence with an acceptable accuracy in the perturbed scenario. Further, to quantitatively compare the steady-state performance, the root mean square (RMS) values, denoted as ${\rm rms}(\cdot)$, of the tracking and estimation errors at the steady state ($40\,\rm sec - 100\,\rm sec$) under different controllers are summarized in Table \ref{tab:table1}. As can be seen, the CI\&IAC shows smallest RMS values of tracking and estimation errors among all three controllers. In summary, the proposed CI\&IAC has an inherent robustness against external disturbances and measurement noises.

%

\begin{table}[hbt!]
	\caption{\label{tab:table1} RMS values of the tracking and estimation errors.}
	\centering
	\begin{threeparttable}
		\begin{tabular}{lccc}
			\hline
			&\multicolumn{3}{c}{RMS value\tnote{a}}\\
			\cline{2-4}
			Method & ${\rm rms}(\bm{q}_{ev})$ & ${\rm rms}(\bm{\omega}_{e})$, $\rm rad/sec$& ${\rm rms}(\tilde{\bm{\theta}})$, $\rm kg \cdot m^{2}$ \\\hline
			CI\&IAC & $4.803\times10^{-4}$     & $9.234\times10^{-4}$ & 0.1433\\
			NCEAC   & $6.165\times10^{-4}$     & 0.0012               & 5.2812\\
			CEAC    & 0.0094                   & 0.0181               & 12.7127\\
			\hline
		\end{tabular}
		\begin{tablenotes}
			\footnotesize
			\item[a] It is taken as the maximum RMS value across all vector components.
		\end{tablenotes}
	\end{threeparttable}
\end{table}

\section{Conclusion} \label{secVI}

This paper addresses the problem of quaternion-based adaptive controller design for anti-unwinding rigid body attitude tracking, in the presence of inertia uncertainties. A composite I\&I adaptive control scheme is proposed, which essentially ensures exponential stability of the resulting closed-loop dynamics under a strictly weak IE condition, and consequently, guarantees exponential convergence of both the output-tracking and parameter estimation errors to zero without causing unwinding. The key ideas behind permitting exponential stability without requiring PE are two-fold: 1) a logarithmic barrier function is used as the attitude error function for unwinding avoidance, along with the tactful establishment of two algebra properties for exponential stability analysis; 2) by virtue of a constructive LTV filter, an LRE extension procedure is proposed for DREM to generate a persistently exciting regressor, based on which a DREM-based learning law is presented for relaxing the dependence of parameter convergence on the PE condition. Saliently, the control algorithm developed preserves all the key beneficial features of the I\&I adaptive control methodology and does not involve any dynamic gains. In addition, the composite learning law is augmented with a power term to achieve the synchronized finite/fixed-time parameter convergence. Finally, simulation results show the effectiveness and superiority of the proposed method. 

\section*{Appendix A \\ Proof of Lemma \ref{Lemma1}}

Two cases are considered to complete the proof. 

Case 1: $q_{e4}\in(0,1]$. For analysis, we introduce an auxiliary variable defined by $h(q_{e4})=(1-q_{e4}^{2})/q_{e4}+\ln q_{e4}^{2}$. Via simple algebraic manipulations, it is shown that
\begin{equation}
\label{eqA1}
\dfrac{\partial h(q_{e4})}{\partial q_{e4}}=-\dfrac{(1-q_{e4})^{2}}{q_{e4}^{2}}\leq0 \tag{A1}
\end{equation}
in the set $q_{e4}\in(0,1]$, indicating that $h(q_{e4})$ is a non-increasing function of $q_{e4}$. Thus, $h(q_{e4})\geq h(1)=0$ holds in this case, which coincides with the result in  \eqref{eq13}. 

Case 2: $q_{e4}\in[-1,0)$. In this case, let us define $h(q_{e4})=-(1-q_{e4}^{2})/q_{e4}+\ln q_{e4}^{2}$. Following a similar reasoning as Case 1, it is not difficult to check that $\partial h(q_{e4})/\partial q_{e4}\geq0$ and hence $h(q_{e4})\geq h(-1)=0$ in the set $q_{e4}\in[-1,0)$, so that  \eqref{eq13} is also obtained. 

Combining the above two cases warrants Lemma \ref{Lemma1}.   $\hfill \blacksquare$

\section*{Appendix B \\ Proof of Lemma \ref{Lemma2}} \label{AppendixB}

Let us first prove that $\underline{\alpha}\|\bm{q}_{ev}\|^{2}=\underline{\alpha}(1-q_{e4}^{2})\leq V_{q}$. To facilitate the analysis, similar to the proof of Lemma \ref{Lemma1}, we here define $h_{1}(q_{e4})=-\alpha\ln(q_{e4}^2)-\underline{\alpha}(1-q_{e4}^2)$. Taking the partial derivative of $h_{1}(q_{e4})$ w.r.t. $q_{e4}$ gives
\begin{equation}
\label{eqB1}
\dfrac{\partial h_{1}(q_{e4})}{\partial q_{e4}}=-\dfrac{2}{q_{e4}}(\alpha-\underline{\alpha} q_{e4}^{2}) \tag{B1}
\end{equation}
since $\underline{\alpha}\leq\alpha$ and $q_{e4}^{2}\leq1$, an intuitive observation reveals that $(\alpha-\underline{\alpha} q_{e4}^{2})\geq0$. With this in mind, one can claim that for $q_{e4}\in(0,1]$, $\partial h_{1}(q_{e4})/\partial q_{e4}\leq0$, while for $q_{e4}\in[-1,0)$, $\partial h_{1}(q_{e4})/\partial q_{e4}\geq0$. From the above, together with the fact that $\lim_{q_{e4}\to0}h_{1}(q_{e4})=+\infty$, it can be concluded that $h_{1}(q_{e4})\geq h(\pm1)=0$. Thus, $\underline{\alpha}(1-q_{e4}^2)\leq-\alpha\ln q_{e4}^2$ always holds for any $\underline{\alpha}\leq1$, so does the inequality $\underline{\alpha}\|\bm{q}_{ev}\|^{2}\leq V_{q}$. 	 

Next we prove that $V_{q}\leq\overline{\alpha}\|\bm{q}_{ev}\|^{2}=\overline{\alpha}(1-q_{e4}^{2})$ holds for any $|q_{e4}|\in[\delta,1)$. To this end, define an auxiliary variable in the set $\delta\leq|q_{e4}|\leq1$ as follows:
\begin{equation}
\label{eqB2}
h_{2}(q_{e4})=\dfrac{-\ln q_{e4}^{2}}{1-q_{e4}^{2}} \tag{B2}
\end{equation}
whose partial derivative w.r.t. $q_{e4}$ is given by
\begin{equation}
\label{eqB3}
\dfrac{\partial h_{2}(q_{e4})}{\partial q_{e4}}=\dfrac{-2(1-q_{e4}^{2})-2q_{e4}^{2}\ln q_{e4}^{2}}{q_{e4}(1-q_{e4}^{2})^{2}} \tag{B3}
\end{equation}
For brevity, we denote by $P(q_{e4})$ the numerator of  \eqref{eqB3}. Taking its partial derivative w.r.t. $q_{e4}$ gives
\begin{equation}
\label{eqB4}
\dfrac{\partial P(q_{e4})}{\partial q_{e4}}=-4q_{e4}\ln q_{e4}^{2} \tag{B4}
\end{equation} 
from which it is not difficult to check that $\partial P(q_{e4})/\partial q_{e4}>0$ for $q_{e4}\in[\delta,1)$ and $\partial P(q_{e4})/\partial q_{e4}<0$ for $q_{e4}\in(-1,-\delta]$. Consequently, the the maximum value of $P(q_{e4})$ takes $\lim_{q_{e4}\to\pm1}P(q_{e4})=0$ in the set $\delta\leq|q_{e4}|<1$, indicating that $P(q_{e4})<0$ for all $|q_{e4}|\in[\delta,1)$. In view of this, from  \eqref{eqB3}, it is clear that $\partial h_{2}(q_{e4})/\partial q_{e4}<0$ for $q_{e4}\in[\delta,1)$ and $\partial h_{2}(q_{e4})/\partial q_{e4}>0$ for $q_{e4}\in(-1,-\delta]$, whereby one can observe that $h_{2}(q_{e4})<-\ln \delta^{2}/(1-\delta^{2})$ for any $|q_{e4}|\in[\delta,1)$. By using simple arithmetic operations, we can conclude that $-\alpha\ln q_{e4}^2<-\alpha(\ln\delta^{2}/(1-\delta^{2}))(1-q_{e4}^{2})$ for any $|q_{e4}|\in[\delta,1)$. As $\overline{\alpha}\geq-\alpha\ln\delta^{2}/(1-\delta^{2})$, it can be further claimed that $-\alpha\ln q_{e4}^2<\overline{\alpha}(1-q_{e4}^{2})$ holds $\forall$ $|q_{e4}|\in[\delta,1)$. In addition, it is noted that $-\alpha\ln q_{e4}^2=\overline{\alpha}(1-q_{e4}^{2})=0$ when $|q_{e4}|=1$. Based on the above argument, we can draw the conclusion that $-\alpha\ln q_{e4}^2\leq\overline{\alpha}(1-q_{e4}^{2})$ holds for any $|q_{e4}|\in[\delta,1]$. 

Furthermore, it can be readily verified that $-\ln \delta^{2}/(1-\delta^{2})>1$ strictly holds for $0<\delta<1$. This directly contributes to the fact that $\bar{\alpha}>\underline{\alpha}$, thus completing the proof. $\hfill \blacksquare$ 




\ifCLASSOPTIONcaptionsoff
  \newpage
\fi



\bibliographystyle{IEEEtran}
\bibliography{refs}
\end{document}